\documentclass{article}

\usepackage{stfloats}
\usepackage{url}
\usepackage{verbatim}
\usepackage{algorithm} 
\usepackage{algorithmic}  
\usepackage[algo2e]{algorithm2e} 
\usepackage{amsmath,amssymb,amsfonts}
\usepackage{algorithm}
\usepackage{graphicx}
\usepackage{textcomp}
\def\BibTeX{{\rm B\kern-.05em{\sc i\kern-.025em b}\kern-.08em
    T\kern-.1667em\lower.7ex\hbox{E}\kern-.125emX}}
    
    \usepackage{lineno}
\usepackage{lettrine}

\usepackage{url}
\usepackage{moreverb}

\usepackage{graphicx}
\usepackage{subcaption}
\usepackage{float}
\usepackage{amsmath}
\usepackage{array}
\usepackage{multirow}
\graphicspath{ {Figures/} }

\usepackage{scalerel}
\usepackage{tikz}
\usetikzlibrary{svg.path}

\definecolor{orcidlogocol}{HTML}{A6CE39}
\tikzset{
  orcidlogo/.pic={
    \fill[orcidlogocol] svg{M256,128c0,70.7-57.3,128-128,128C57.3,256,0,198.7,0,128C0,57.3,57.3,0,128,0C198.7,0,256,57.3,256,128z};
    \fill[white] svg{M86.3,186.2H70.9V79.1h15.4v48.4V186.2z}
                 svg{M108.9,79.1h41.6c39.6,0,57,28.3,57,53.6c0,27.5-21.5,53.6-56.8,53.6h-41.8V79.1z M124.3,172.4h24.5c34.9,0,42.9-26.5,42.9-39.7c0-21.5-13.7-39.7-43.7-39.7h-23.7V172.4z}
                 svg{M88.7,56.8c0,5.5-4.5,10.1-10.1,10.1c-5.6,0-10.1-4.6-10.1-10.1c0-5.6,4.5-10.1,10.1-10.1C84.2,46.7,88.7,51.3,88.7,56.8z};
  }
}

\newcommand\orcidicon[1]{\href{https://orcid.org/#1}{\mbox{\scalerel*{
\begin{tikzpicture}[yscale=-1,transform shape]
\pic{orcidlogo};
\end{tikzpicture}
}{|}}}}

\usepackage[nonatbib,final]{neurips_2023_ml4ad}

\usepackage[utf8]{inputenc} 
\usepackage[T1]{fontenc}    
\usepackage[hidelinks]{hyperref}       
\usepackage{url}            
\usepackage{booktabs}       
\usepackage{amsfonts}       
\usepackage{nicefrac}       
\usepackage{microtype}      
\usepackage{xcolor}         

\title{Development and Assessment of Autonomous Vehicles in Both Fully Automated and Mixed Traffic Conditions}

\author{%
Ahmed Abdelrahman\\
  University of Central Florida\\
  \texttt{ahmed.abdelrahman@ucf.edu} \\
}

\begin{document}



\maketitle

\begin{abstract}
Autonomous Vehicle (AV) technology is advancing rapidly, promising a significant shift in road transportation safety and potentially resolving various complex transportation issues. With the increasing deployment of AVs by various companies, questions emerge about how AVs interact with each other and with human drivers, especially when AVs are prevalent on the roads. Ensuring cooperative interaction between AVs and between AVs and human drivers is critical, though there are concerns about possible negative competitive behaviors. This paper presents a multi-stage approach, starting with the development of a single AV and progressing to connected AVs, incorporating sharing and caring V2V communication strategy to enhance mutual coordination. A survey is conducted to validate the driving performance of the AV and will be utilized for a mixed traffic case study, which focuses on how the human drivers will react to the AV driving alongside them on the same road. Results show that using deep reinforcement learning, the AV acquired driving behavior that reached human driving performance. The adoption of sharing and caring based V2V communication within AV networks enhances their driving behavior, aids in more effective action planning, and promotes collaborative behavior amongst the AVs. The survey shows that safety in mixed traffic cannot be guaranteed, as we cannot control human ego-driven actions if they decide to compete with AV. Consequently, this paper advocates for enhanced research into the safe incorporation of AVs on public roads.
\end{abstract}



\section{Introduction}
The advent of automation heralds a new era with Autonomous Vehicles (AVs) playing a central role in this transition. As a cornerstone of modern technological transformation, these vehicles have garnered attention from various industries and research groups. They promise to revolutionize Intelligent Transportation Systems (ITS) by bolstering safety mechanisms. Defined by the J3016 standard \cite{TaxonomyAD}, AVs can range from level 1, offering basic driver assistance, to level 5, which boasts complete autonomy. Crucially, vehicles utilize sensors, like cameras and radars, to gauge and interact with their environment. However, each sensor type, much like human vision, comes with its own limitations, specifically in terms of field of view (FOV).\\
This segue into the human element brings us to a pressing concern in transportation: human error. In fact, driver inattentiveness, a subset of human errors, has been linked to approximately 80\% of crashes and 65\% of near-crashes, according to a study from the National Highway Traffic Safety Administration \cite{dingus2006100}. To mitigate such issues, the concept of cooperative autonomous driving emerges as a beacon of hope. By leveraging vehicle-to-vehicle (V2V) communication, AVs can apprise other vehicles of their position, even beyond traditional FOVs. This becomes especially vital in places like intersections, where structures often obscure visibility. A sobering statistic to consider is that, in 2008, nearly 40\% of an estimated 5,811,000 collisions in the US were attributed to intersection-related incidents \cite{choi2010crash}. Such data undeniably emphasize the pressing need for continued innovation in ITS technologies.
Building on this, the emergence of V2V, a communication modality amongst vehicles, further reinforces the ethos of collaboration in vehicular systems. Beyond just enhancing safety, V2V offers the potential to optimize road capacity, cut down on fuel consumption, and minimize harmful emissions \cite{outay2019v2v,van2017robust}. To put this into perspective, transportation, as a sector in the European Union, was identified as a source for roughly 25\% of all CO2 emissions, as per a report by the European Environment Agency (EEA) \cite{ferreira2011impact}.\\
Steering the discussion back to AVs, it's evident that their successful integration hinges on their precise motion control. While traditional vehicle motion control strategies, such as backstepping and model predictive control (MPC), have had their merits \cite{shalaby2019design}, they also come with inherent challenges. For one, they necessitate detailed vehicle models, complicating systems and demanding heavy computational resources. This is where deep Reinforcement Learning (RL) makes its mark \cite{chu2019model}. A study by Farag et al., for instance, showcased the edge of deep RL over MPC in terms of efficiency and control accuracy in vehicle group formations \cite{faragreinforcement}. Moreover, several studies have spotlighted the potential of model-free deep RL in vehicle motion control \cite{jaritz2018end, cai2020high}. Some even argue that with the aid of deep RL, AV performance can transcend human driving skills \cite{fuchs2021super}.\\
The integration of multiple AVs opens up numerous possibilities and applications, ultimately simplifying our lives. However, this stage also brings about various concerns and fears \cite{yang2021safety}. How will AVs from different manufacturers behave with each other, and how human drivers will react to the AVs around them? Will they cooperate with each other, or a lot of competition will occur between them?

After examining the literature in addition to the survey conducted on applying deep RL for AV motion control \cite{aradi2020survey}, there exist some research gaps listed as follows:

\begin{itemize}
    \item Examine empirically the implications that arise when the AVs are not interconnected.
    \item Mixed traffic has not been extensively addressed, which is an inevitable transition to fully automated traffic \cite{ozioko2022road}.
    \item In some studies low physics simulators are used, where there is a huge gap between the simulation and real-life work.

\end{itemize}

The main contribution of this paper involves three key aspects. First, it focuses on the development of a single AV that closely emulates human driving capabilities. Second, it scales up from a solitary AV to multiple AVs, exploring multiple communication networks among them. This focuses on examining the influence of connectivity among AVs and contrasting fully automated traffic with mixed traffic environments, where AVs and human drivers coexist.
The simulation work is conducted on Unity, a powerful 3D physics game engine. Within this urban environment created in Unity, different test scenarios are executed while incorporating human-in-the-loop in some of the experiments to have nuanced insights. These tests assess the performance of a single AV in terms of driving behavior, followed by evaluations of interconnected AVs, and additionally, investigate mixed traffic scenarios. This approach provides valuable empirical insights, especially in contexts where AV and human-driven vehicle (HDV) interact.\\
The paper is organized as follows: In Section 2, the development of connected AVs is presented. Section 3 presents the outcomes of the experiments conducted and underscores the key takeaways from these results. Finally, Section 4 concludes the paper, summarizing the key findings and contributions.

\section{Methodology}\label{sec3}

This section dives into the development of Cooperative Autonomous Driving (CAD) between two Connected AVs driving on the same road track powered by deep RL to train the agents to learn driving behavior that can reach human driving performance. Figure 1 shows the Markov Decision Process (MDP) of the CAD system which describes the dynamic interaction between the agent (AV) and the environment (Road Track) with a loop of actions, states, and rewards. The agent takes an action and this action puts the agent in a new state with respect to the dynamic environment and results in feedback from the environment to the agent in the form of rewards which can be a positive reward, good feedback, and thus good taken action, or a penalty, bad feedback, and thus bad taken action \cite{sutton2018reinforcement}.
This system is considered a multi-agent system where agents interact with each other in the same environment. As visualized in Figure 1, the blue vehicle is subjected to train driving on the right lane and the red vehicle is subjected to train driving on the left lane. Both vehicles have the same main objectives which are driving safely (avoid crashes through the road track), and efficiently (reach the finish line in the shortest time).

\begin{figure}
     \centering
     \begin{subfigure}[b]{0.5\textwidth}
         \centering
         \includegraphics[width=\textwidth]{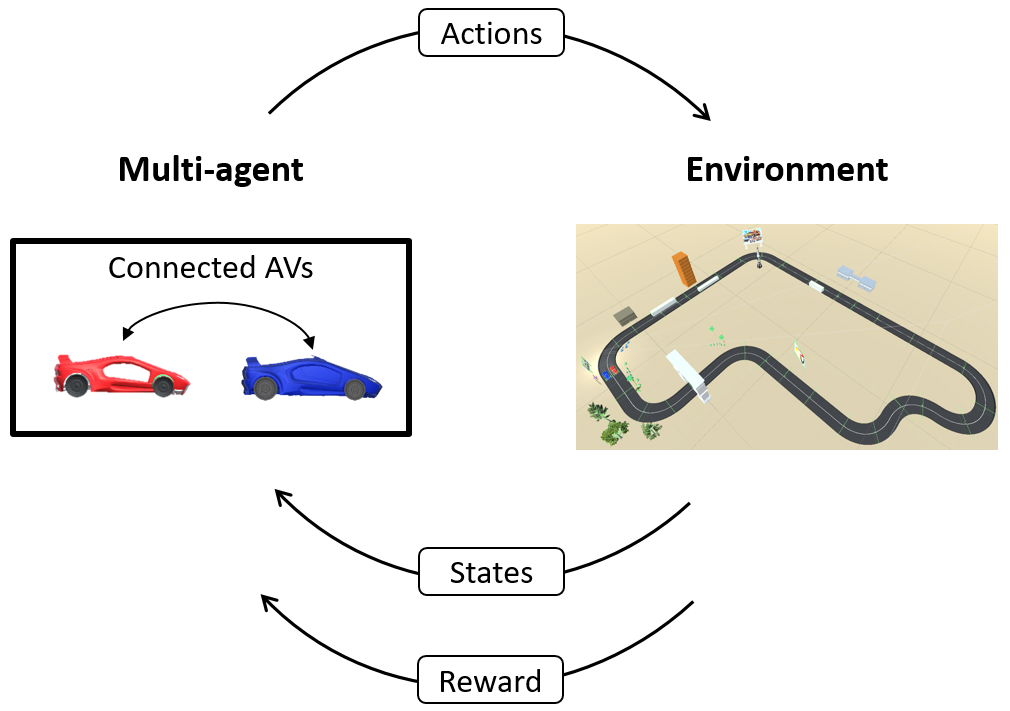}
         \label{fig:y equals x}
     \end{subfigure}
     \hfill
     \begin{subfigure}[b]{0.45\textwidth}
         \centering
         \includegraphics[width=\textwidth]{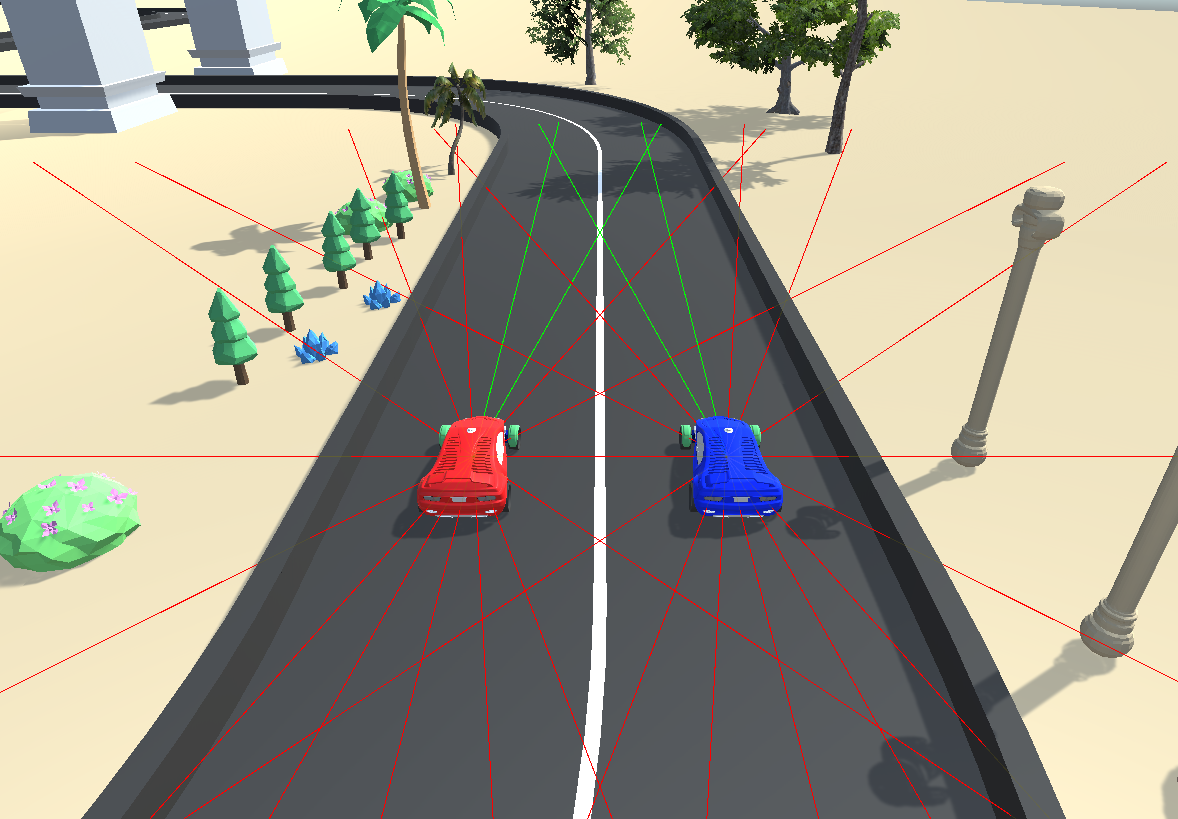}
         \label{fig:five over x}
     \end{subfigure}
     
        \caption{MDP for the two connected vehicles}
        \label{fig:three graphs}
\end{figure}

\subsection{Multi-Agent}
This is a multi-agent system where agents interact with each other in the same environment to achieve the maximum accumulative reward through training with deep RL. One vehicle (blue vehicle) has to drive in the right lane and the other one (red vehicle) in the left lane unless they should do lane changing to avoid any dangerous situation like crashing obstacles.
Vehicles are trained using Proximal Policy Optimization (PPO) based end-to-end deep RL to learn a robust cooperative driving behavior. PPO has proven its supremacy in many learnable tasks outperforming other RL algorithms while being more empirical for having better sample complexity \cite{juliani2018unity, ye2020automated}. Most deep RL and PPO hyperparameters are set to the default values supported by the ML-Agents toolkit or adopted from other studies' recommendations \cite{juliani2018unity, gupta2021embodied}.\\
The Agent is based on the SkyCar vehicle model from Unity's standard assets (2018.4) designed to be small for easier agent training, with a mass of 1000 kg, top speed of 40 km/h, and maximum steering angle of 25 degrees.
To navigate its path and avoid collisions, the vehicle uses inputs from its perception system, consisting of its ego velocity and 16 raycast sensors, visualized in Figure 1, which function like laser distance sensors measuring up to 50 meters.

\subsection{Environment}
This work is conducted in an urban environment with certain rules, such as each vehicle has to drive in the assigned lane while avoiding any crashes. Track structure, checkpoints, and obstacles form the environment, as shown in Figure 2, which the agents will interact with. Environment corresponds to agents' actions and sends back positive or negative rewards and presents new situations to the agents. With these interactions, vehicles learn how to drive autonomously on the road track while maintaining safe driving behavior.
The road track is designed to incrementally increase driving complexity, starting with straight-ways and advancing to multiple turns and obstacles. Based on the American Association of State Highway and Transportation Officials (AASHTO) and The Federal Highway Administration (FHWA)  \cite{aashto2001policy, FHWA_lane}, the lane width is set at 3.5 meters, reflecting urban U.S. road standards, with friction coefficients of 0.3 and 0.25. The track's borders facilitate lane detection, and driving without collisions emulates real-life scenarios with structural borders. To guide vehicles and ensure they remain in their assigned lanes, invisible checkpoints offer reward-based guidance, emphasizing that changing lanes to avoid an obstacle incurs a lesser penalty than crashing. The track also includes static obstacles like rigid boxes, representing stationary roadblocks, and dynamic obstacles, such as another moving vehicle on the same track.

\begin{figure}
    \centering
    \includegraphics[width=0.9\linewidth]{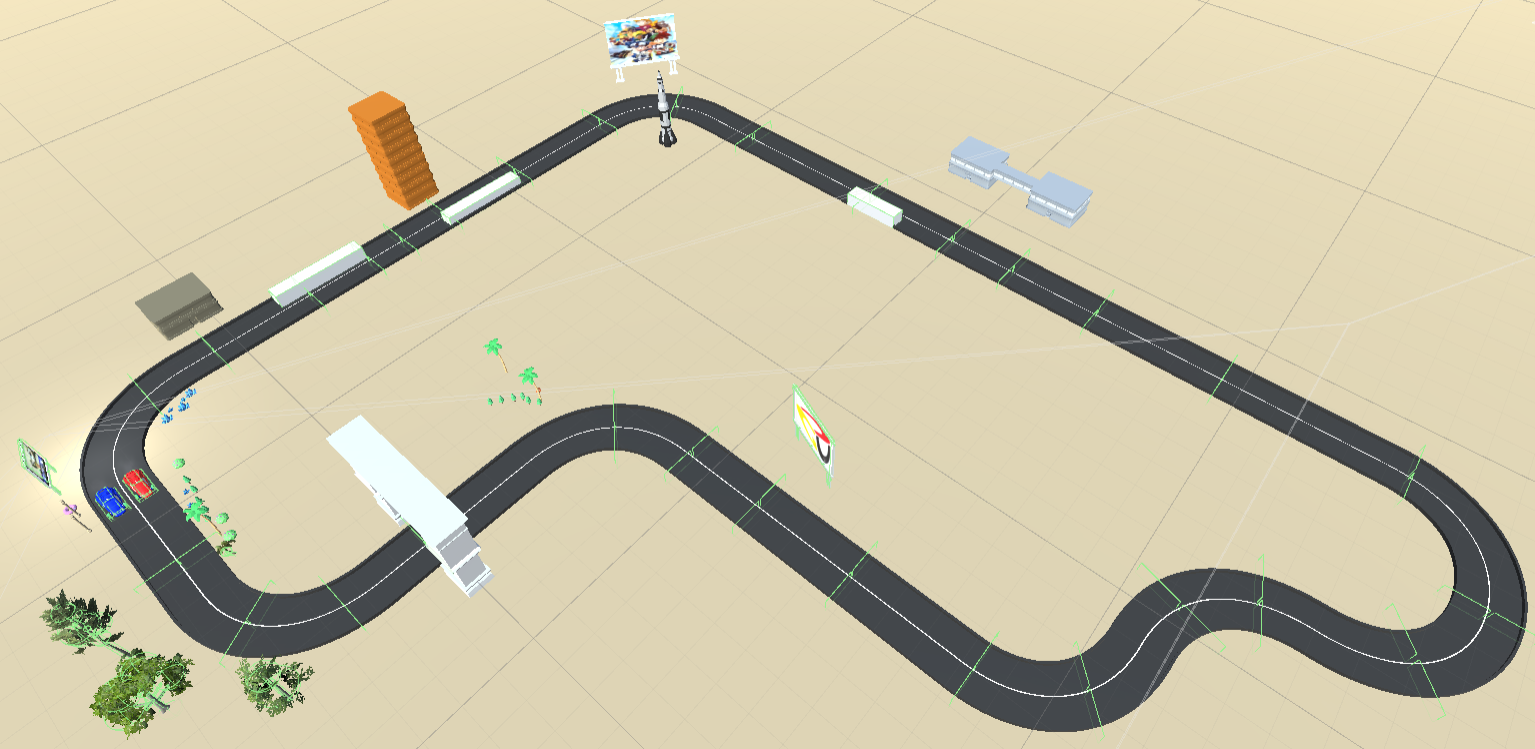}
    \caption{Urban training environment for connected AVs}
\end{figure}


\subsection{MARL Model Training for Connected AVs}
CAD for Connected AVs is a Multi-Agent Reinforcement Learning (MARL) problem that requires extensive training processes to acquire smooth driving like humans in addition to learning cooperation between each other. Figure 3 visualizes how this multi-stage system proceeded for both of the two vehicles. Each of the first two stages can be defined as Single-Agent Reinforcement Learning (SARL) problem where each vehicle is trained solely on the road track, then the final two stages are MARL problems where both vehicles drive on the road track and interact with each other and share their perception data. Throughout this whole multi-stage training process, both vehicles have the same model to have a more intelligent model that works for multiple styles of driving in addition to learning CAD behavior. This is applicable through transfer learning of the same model \cite{tan2018survey}. To inform the model which lane is assigned to drive on, one of the inputs to the model is 0 for driving on the right lane and 1 for the left lane.

\begin{figure}
    \centering
    \includegraphics[width=1\linewidth]{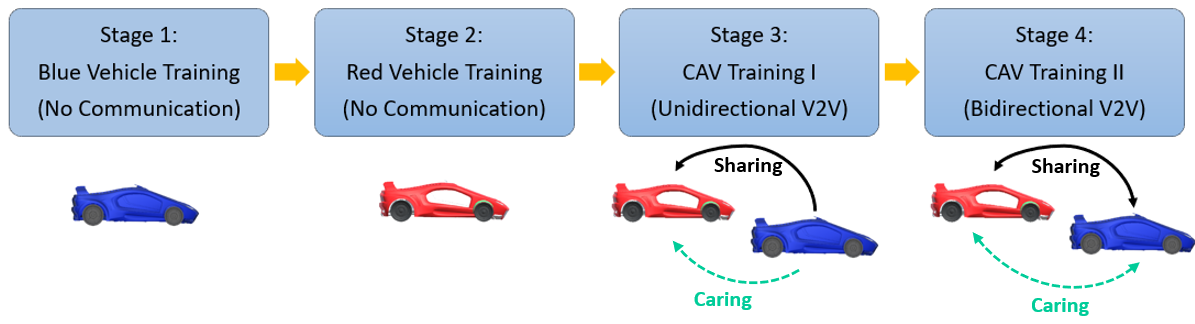} 
    \caption{CAD model training stages}
\end{figure}

\subsubsection{First and Second Stages}
The first two stages can be considered the first level of training since this is the first level of training to drive for both red and blue vehicles. Learning the basics of driving is essential like how to drive in the right or left lane of the track without crashes with track borders or the present obstacles. In the first stage, the blue vehicle trains driving in the right lane. In the second stage, the red vehicle trains driving in the left lane. Both vehicles do not share any data between them in these two stages.

\subsubsection{Third Stage $|$ Unidirectional Communication Topology}
In the third stage, only one vehicle is connected with the other vehicle and receives shared information from it through V2V as shown in Figure 3 (stage 3). The red vehicle is the only vehicle that is training in this stage while the blue vehicle is driving based on the trained model from stage 1. The shared information consists of the blue vehicle's 16 raycast sensor readings, velocity, and position.

\subsubsection{Fourth Stage $|$ Bidirectional Communication Topology}
This is the final stage where the two vehicles are connected through V2V and share each other's perception data and position through bidirectional V2V as shown in Figure 3 (stage 4). But model training this time is applied only on the blue vehicle while the red vehicle is driving based on the trained model from stage 3.\\
\textbf{Sharing and Caring based V2V:}
During model validation, we observed how the two AVs cautiously interacted when encountering a lane-blocking obstacle. They seemed hesitant, balancing the risk of collision against potential penalties for speed reduction or lane changes. This mirrored human drivers' instincts to assist yet maintain self-interest. To resolve this, we introduced a "sharing and caring" concept between them. It pairs AVs into supportive units that exchange data and provide mutual aid. The 'sharing' aspect is the V2V perception data exchange, while 'caring' refers to a reward signal, where a vehicle earns a positive reward when its partner successfully reaches the destination (finish line).
Although this concept also has been applied to unidirectional V2V, it is most effective and fully realized in bidirectional V2V communication.
After this final stage, the two connected AVs learn how to drive together cooperatively in the dynamic environment and gain the maximum possible accumulative reward while sharing and caring.

\subsubsection{RL Reward Signals}
Defining the reward values in the case of MARL is a little bit more complex than in the case of SARL where just a single vehicle trains how to drive solely on the track. Some unexpected behavior occurred while training the two vehicles together, sometimes vehicles are afraid of driving or decide that the best action is to not drive and this is because of applying high penalties for crashing. Vehicles at the beginning of training learn the basics of driving and for sure they would commit many aggressive actions and crashes that yield to gaining penalties. Another unexpected behavior is that the two vehicles are very ego about their accumulative rewards and while training the agents learned a very aggressive and malicious behavior when they realized that whenever the vehicle crashes it goes back to the start line; the vehicle drives very close to the other vehicle and pushes it towards the track border to crash and goes back to the start line, so now it can move freely on the track without any interaction with the other vehicle, and sometimes both vehicles, because of these aggressive maneuvers, crash each other. To solve this aggressive driving behavior, the model is subjected to the proposed four-stage training process, and the bidirectional communication topology is applied between connected AVs to boost coherence between them.
The structuring of reward signals is based on the linguistic criteria outlined below, while Table 1 presents the numerical values for these signals:
\begin{itemize}
\item The vehicle will crash if it touches the track border or any object.
\item The vehicle should reach the finish line of the track as soon as possible.
\item It is better to not frequently steer and drive smoothly to ensure comfortable driving.
\item When connected, engage in the practice of mutual sharing and caring.
\end{itemize}

\begin{table}
\centering
\caption{Reward signals for MARL problem (CAD)} 
\resizebox{13.8cm}{!}{
\renewcommand{\arraystretch}{1.5}
\Large
\begin{tabular} {|c||c|c|c|c|c|c|c|c|} 
 \hline
 \textbf{Event} & \textbf{lap time} & \textbf{Subjected lane CKPT} & \textbf{other lane CKPT} & \textbf{Finish line} & \textbf{Hit obstacle} & \textbf{Crash} & \textbf{Drive smooth} & \textbf{Caring}\\
 \hline
 \textbf{Reward} & \textbf{-time} & \textbf{+1} & \textbf{-2} & \textbf{+100} & \textbf{-5} & \textbf{-10} & \textbf{+0.1 per timestamp} & \textbf{+100}\\
 \hline
\end{tabular} } 
\end{table}

\subsubsection{Deep Neural Network Structure}
The deep Neural Network (DNN) consists of 5 layers of shape \{37,256,256,256,2\}. Figure 4 visualizes the Network, the input layer that takes the ego perception, the assigned lane indicator, and the connected vehicle's perception. Then 3 fully connected hidden layers with 256 neurons each, and the output layer consists of 2 neurons which are the steering and throttle of each vehicle. The V2V perception input data is off (set to zeros) for the first two training stages and will be used when the V2V communication topologies are activated between the two AVs in the third and fourth training stages.

\begin{figure}
    \centering
    \includegraphics[scale=0.2]{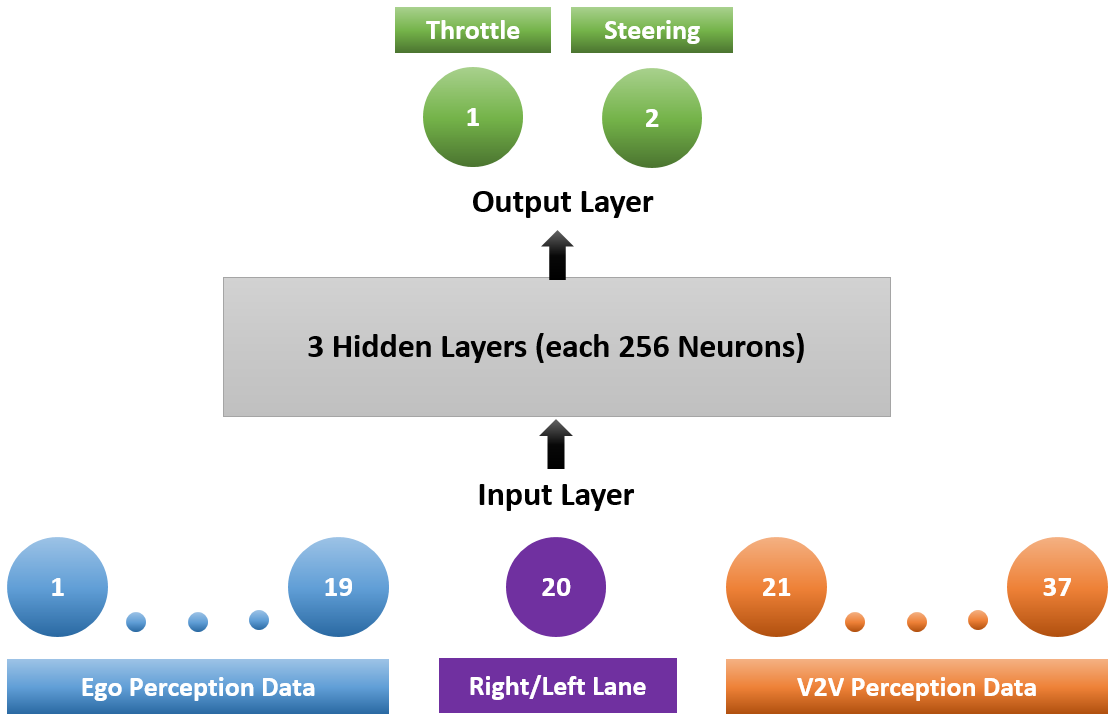}
    \caption{Deep neural network layers for CAD}
\end{figure}
\section{Results and Discussion}\label{sec5}
\subsection{Stage 1 $|$ Single Autonomous Vehicle (SAV) vs. Human Driven Vehicle (HDV)}
Evaluating the results of the SAV, trained using deep RL, involves comparing its performance with HDV on the same road track on Unity 3D simulator.
To validate the trained deep RL model for SAV driving performance, the model is tested in the same training environment but after changing its start position and obstacle locations on the track to check model robustness and make sure no overfitting to the start position or obstacle locations. The trained deep RL driving model for the AV has successfully reached the finish line of the 10 laps without any crashes, lap time of each lab is shown in Table A.1.

\textbf{Survey Experiment 1 Results:} Participants play 5 times on the same track of the AV but each one drives alone on the right lane of the track. This experiment is to measure the difference in the average lap time between the HDV for each participant and the AV. 5 laps were recorded for the AV and 5 laps for each participant. The average lap time of the AV is 59.67 seconds which will be compared with the HDVs average lap time. The results of each lap are shown in Table A.3 and Table 2 shows the summary of the conducted solo race between the HDVs and the AV driving in the testing urban environment.

\begin{table}
\centering
\caption{Statistics $|$ HDV vs AV in mean lap time}
\renewcommand{\arraystretch}{1.2}
\begin{tabular} {|c||c|c|} 
 \hline
 \textbf{Playing Level} & \textbf{HDV Wins} & \textbf{AV Wins}\\
\hline
Pro Level (5 persons) & 2 (40\%) & 3 (60\%)\\ 
\hline
Intermediate Level (4 persons) & 0 (0\%) & 4 (100\%)\\
\hline
Beginner level (2 persons) & 0 (0\%) & 2 (100\%)\\
\hline
\end{tabular}
\end{table}

\begin{figure}
\centering
\begin{subfigure}{.45\textwidth}
  \centering
  \includegraphics[width=\linewidth]{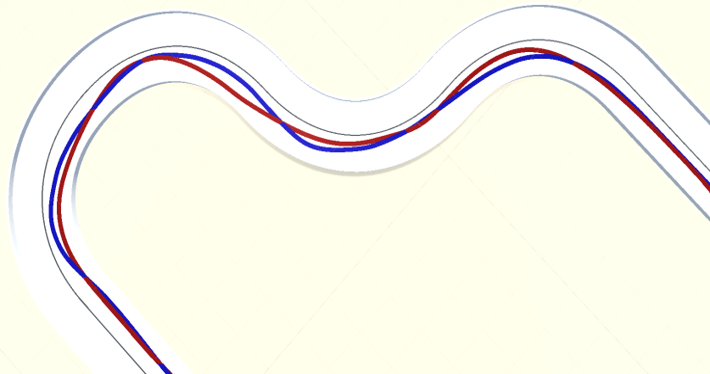}  
  \caption{Trajectory tracking in turns}
\end{subfigure}
\hfill
\begin{subfigure}{.40\textwidth}
  \centering
  \includegraphics[width=\linewidth]{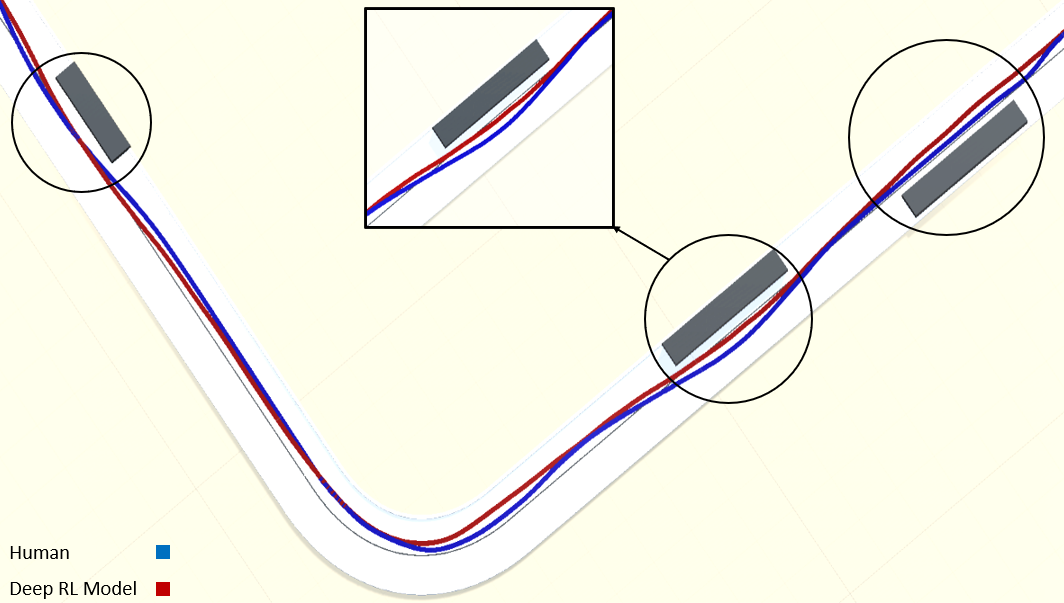}  
  \caption{Trajectory tracking through the obstacles}
\end{subfigure}
\caption{HDV vs AV trajectory tracking}
\label{fig:fig}
\end{figure}

\textbf{Discusion:} The trajectory tracking of both the best-performing HDV from the 11 participants and the AV driving on the right lane is recorded to analyze their driving performance. Figures 5(a) and 5(b) reveal that the driving behavior of the best HDV (blue line) and the AV (red line) is nearly identical, but the AV demonstrates greater adherence to the assigned lane restriction. Moreover, in obstacle avoidance, the AV exhibits higher precision and efficiency.\\
Based on the numerical results from the first experiment, which involved a race between the participants and the deep RL AV, it is concluded that the trained deep RL model has reached quite similar human driving performance and sometimes exceeds it reaching super-human performance.
SAV has demonstrated superb driving performance in a challenging urban environment, featuring a track with hard turns and multiple large obstacles.

\subsection{Stage 2 $|$ AVs with no V2V}

\textbf{Testing the red AV on the track:} To assess the driving performance of the trained deep RL model for the red vehicle, which operates in the left lane, the model is tested on the same training track. However, the start position and obstacle locations on the track are altered to ensure robustness and ensure no overfitting to specific conditions. The results of 10 laps on the track, as recorded in Table A.2, demonstrate good driving performance with zero crashes in all laps. This reinforces the model's reliability and its capacity to navigate the track effectively under different conditions.

\begin{table}
\centering
\caption{Statistics $|$ Numerical results of stages 2, 3, and 4}
\renewcommand{\arraystretch}{1.2}
\begin{tabular} {|c||c|c|c|} 
 \hline
 \textbf{Subject} & \textbf{Stage 2}  & \textbf{Stage 3}  & \textbf{Stage 4}\\
\hline
Total Number of Crashed laps: & 13 & 1 & 0\\ 
\hline
Accidents Percentage: & 65\% & 10\% & 0\%\\
\hline
Safe Driving Percentage: & 35\% & 90\% & 100\%\\
\hline
Blue AV Win Race Percentage: & 14\% & 0\% & 0\%\\
\hline
Red AV Win Race Percentage: & 86\% & 100\% & 100\%\\
\hline
\end{tabular}
\end{table}

\textbf{Two AVs without V2V} To verify the importance of V2V communication between AVs, 20 laps are recorded of the blue AV vs. the red AV with no V2V.
The results in Table 3 (Stage 2) indicate a high accident rate of 65\%, highlighting a lack of cooperation between the AVs. Each AV focused solely on maximizing its individual cumulative results, resulting in crashes and hostile interactions between them. During training, the red AV observed that whenever it collided with the track borders, it would be sent back to the start line and resume driving on the track. Exploiting this knowledge, the red AV intentionally steered towards the blue AV, forcing the other AV to collide with the track border. This malicious action proved advantageous for the red AV, as crashing with another AV incurred a great penalty. Consequently, the blue AV returned to the start line, while the red AV continued its uninterrupted journey in the left lane without facing any harassment from the blue AV.

\subsection{Stage 3 $|$ Unidirectional V2V Results}

\textbf{Testing Unidirectional V2V:} To capture the influence of Unidirectional V2V between connected AVs, 10 laps for the blue AV vs. the red AV are conducted while only the red AV receives the perception data from the blue AV.
In the laps records shown in Table 3 (Stage 3), the percentage of accidents decreased significantly to just 10\% after applying the unidirectional communication topology between the two AVs. When one (red AV) of the two vehicles knows the other vehicle's perception data and its position on the track, it is able to locate it and avoid any crashes between them as possible. However, the problem of accidents is not solved completely because the blue vehicle, the one that does not receive perception data from the red vehicle, cannot see the other vehicle. This limitation is attributed to the exclusive use of laser distance sensors for observing the environment, occasionally leading the blue vehicle to take actions blindly, resulting in hazardous events.

\subsection{Stage 4 $|$ Bidirectional V2V Results}

\textbf{Testing Bidirectional V2V}: Results summary of the blue AV vs. red AV for 10 laps based on Bidirectional V2V are shown in Table 3 (Stage 4). 
When the AVs can mutually observe each other's perception, care about each other's rewards, and compensate in case of any compromises, they achieve a crash-free rate. By driving together, they have established a collaborative-like agreement that allows the AV in the left lane to take the lead. This cooperative approach facilitates smoother driving on the road track and prevents collisions between them, especially during encounters with sudden obstacles or challenging turns.

\subsection{Mixed Traffic (HDV and AV) Results}
\subsubsection{Survey Experiment 2}
Participants play 5 times on the same testing road track, but this time both will drive together until they reach the finish line. The AV drives in the left lane of the track, receiving the HDV perception data and position, and the HDV drives in the right lane.
Experiment 2 investigates the driving behavior of the AV and the HDV. The study explores whether humans would act egoistically towards one another or if the AV would be the weaker, allowing HDV the chance to win. Conversely, it examines the possibility of the AV being egoistic and engaging in aggressive competition for victory.
Table 4 presents a summary of the laps conducted in the mixed traffic scenario. The results reveal that the AV tends to adopt a conservative approach, prioritizing cooperation with the HDV. When successful, this cooperation allows both vehicles to reach the finish line, with occasional instances where the AV or the HDV crosses the finish line first. However, on average, the AV outperforms the HDVs in reaching the finish line faster, thanks to its superior driving performance on the track.
In situations where human drivers adopt aggressive driving behavior and engage in careless maneuvers against the AV, the AV may need to slow down and in some cases intentionally hit the track border to return to the start line, thereby avoiding potential crashes between the two vehicles. This decision is motivated by the higher penalty associated with vehicle accidents compared to merely hitting the track border.

\subsubsection{Results Analysis}

\begin{table}
\centering
\caption{Mixed traffic statistics $|$ HDV vs deep RL AV}
\renewcommand{\arraystretch}{1.2}
\begin{tabular} {|c||c|c|} 
 \hline
 \textbf{Playing Level} & \textbf{Successful Cooperation} & \textbf{Failed Cooperation}\\
\hline
Pro (5) & 22 laps (88\%) & 3 laps (12\%)\\
\hline
Intermediate (4) & 13 laps (65\%) & 7 laps (35\%)\\
\hline
Beginner (2) & 6 laps (60\%) & 4 laps (40\%)\\
\hline
\end{tabular}
\end{table}
In the observations from the mixed traffic experiment, it was noted that many human drivers prioritized their own performance over cooperation. As the driving complexity escalated, human drivers paid more attention to the actions of other vehicles to prevent accidents. Some drivers disregarded lane restrictions in a bid to reach the finish line quicker, despite being informed of these rules before the race. The AV demonstrated proficiency in navigating turns and evading obstacles and was also adept at responding to and avoiding unforeseen actions from the HDV.
The results and collected observations indicate that dealing with mixed traffic poses challenges for human drivers. It requires professional drivers and a conservative driving approach to avoid aggressive maneuvers that could lead to accidents. Consequently, cooperation between AVs and human drivers alone may not be the most effective approach to achieving higher levels of safe driving. Advanced technologies facilitating seamless communication and interactions between HDVs and AVs are necessary.

\subsubsection{Survey $|$ Participants' Feedback}

According to the feedback from participants after engaging in the car game with the AV, the AV was evaluated to have achieved the same or even surpassed human performance. This feedback accounted for a significant percentage (approximately 91\%), reflecting the remarkably good and smooth driving performance of the deep RL model for the AV.

\subsection{CAD and Mixed Traffic Results Summary}
The performance of CAD through V2V communication in a fully automated environment surpasses that in mixed traffic in terms of road safety. No guarantee that all human drivers will be able to deal with the AVs on the roads.
Figure 6 presents a summary of the results in both fully automated traffic and mixed traffic. CAD reached 90\% and 100\% safety in driving on the test road track through unidirectional and bidirectional communication topologies, respectively. This reinforces the importance of the previously discussed problem formulation, which highlights the concerns surrounding mixed traffic, especially in hazardous areas. The power of connected vehicles and cooperative driving can enable higher levels of safety to be achieved.

\begin{figure}
    \centering
    \includegraphics[width=.75\linewidth]{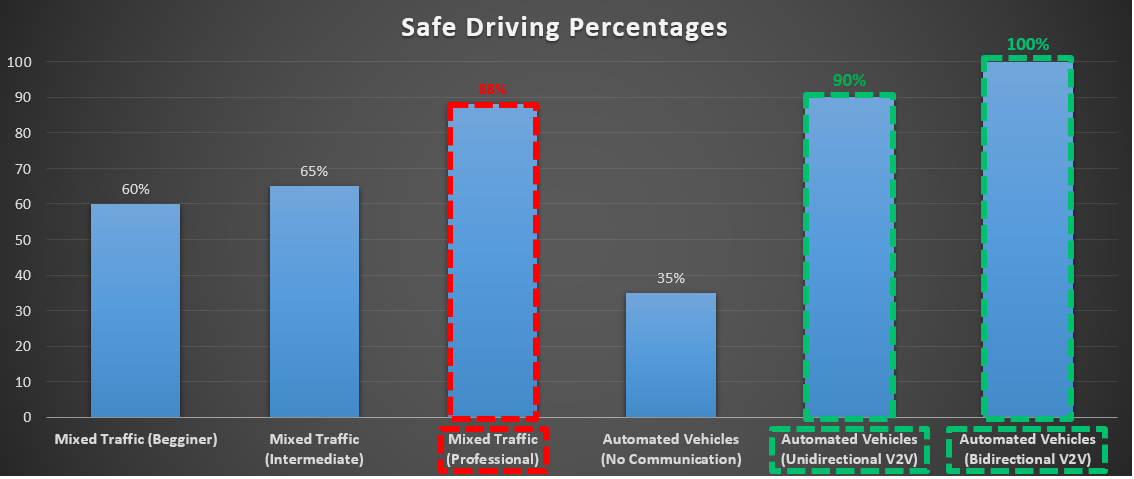}
    \caption{Summary of all experimental results}
\end{figure}

\subsection{Limitations}
The limitations of this experimental study are as follows:
\begin{itemize}
    \item While the use of a computer game offers a unique perspective on studying human driving performance, transitioning to a comprehensive more engaging simulator might provide more realistic insights.
    \item Incorporating components of urban traffic, such as intersections or vulnerable road users, might provide a more holistic view of the challenges in urban driving scenarios.
\end{itemize}

\subsection{Visualized Results}
The following drive links are provided to showcase visual depictions of the experiments conducted, as detailed below:

\begin{enumerate}
    \item Malicious action between AVs in the absence of any cooperation between them, as discussed previously (\url{https://tinyurl.com/ymmtbb8k})
    \item Failed cooperation between the HDV (blue vehicle) and the AV, because of the aggressive maneuver of the HDV (\url{https://tinyurl.com/yr94j46m})
    \item Successful cooperation between the HDV (blue vehicle) and the AV (\url{https://tinyurl.com/yt27ljdx})
    \item Bidirectional communication topology between AVs enhanced their driving behavior while driving next to each other and achieved zero accidents (\url{https://tinyurl.com/yowyyhd3})
\end{enumerate}

\section{Conclusion and Future Recommendations}\label{sec6}
The research examined cooperative autonomous driving, progressing from a single AV to connected AVs using the Unity 3D game engine. This platform facilitated human-in-the-loop experiments, providing a dynamic and interactive environment. The initial phase highlighted the superior performance of a solo AV in navigating complex routes, at times outperforming HDV. Following this, the research focused on the development of connected AVs, emphasizing the impact of the innovative sharing and caring based V2V concept among AVs, which significantly enhanced their cooperative behavior. While 88\% of professional drivers successfully cooperated with AVs, intermediate and beginner drivers showed cooperation rates of just 65\% and 60\%. However, when looking at connected AVs, they exhibited 90\% and 100\% safety rates, depending on their communication type. Overall, fully integrated AVs surpassed mixed traffic in safety and efficient road utilization.\\
For further research, it's recommended to incorporate more AVs in studies, delve deeper into their cooperative potential, and expand mixed traffic studies to enhance our understanding and minimize conflicts between humans and AVs. To address the challenges of mixed traffic with HDVs and AVs, the study advocates for collaboration between AV manufacturers and transportation departments. This would involve enhancing infrastructure and setting up specialized driving academies to train human drivers, ensuring they can safely and efficiently interact with AVs, hence building trust and safety in densely mixed traffic environments.

\newpage
\appendix
\section{Supplementary Material}

\setcounter{figure}{0}
\setcounter{table}{0}
\renewcommand\thefigure{\thesection.\arabic{figure}}
\renewcommand\thetable{\thesection.\arabic{table}}

\subsection{Vehicle (agent) Design}
The Agent is the dynamic model of the SkyCar vehicle model from the standard assets of Unity (2018.4)\cite{UnitySA}. Figure A.1(a) shows the dimensions of the vehicle and it is considered to be a small vehicle to ease training on the agent. To enable physical interactions between objects in Unity, each object should have colliders \cite{UnityC}. Colliders are invisible components that define the shape of a game object for the purposes of physical collisions and they do not need to have the same exact shape as the game object. A rough approximation, as shown in Figure A.1(b), of the mesh is often more efficient and indistinguishable in the gameplay. Additionally, the Wheel Collider is a special collider for grounded vehicles with parameters, mass is 20 kg per wheel, the radius is 0.335 meters, the forward extreme slip is 0.4, and the sideways extreme slip is 0.2. It has built-in collision detection, wheel physics, and the slip-based tire friction model \cite{UnityWC}. 
It can be used for objects other than wheels, but it is specifically designed for vehicles with wheels.\\

\begin{figure}[ht]
     \centering
     \begin{subfigure}[b]{0.45\textwidth}
         \centering
         \includegraphics[width=\textwidth]{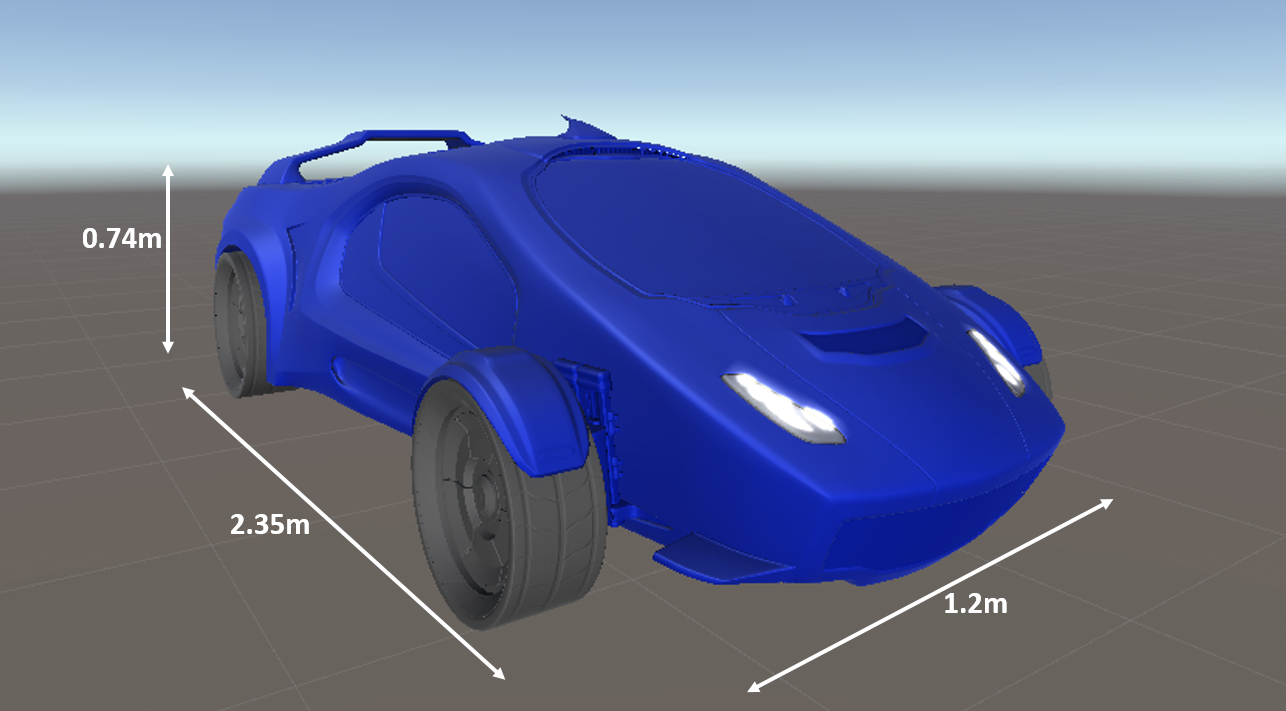}
         \caption{Vehicle's dimensions}
     \end{subfigure}
     \hfill
     \begin{subfigure}[b]{0.45\textwidth}
         \centering
         \includegraphics[width=\textwidth]{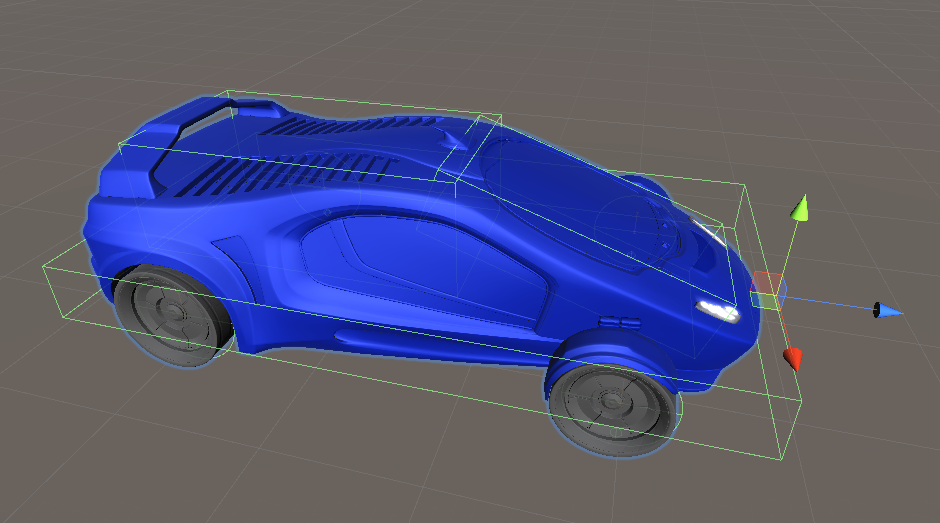}
         \caption{Colliders}
     \end{subfigure}
        \caption{Vehicle design, dimensions, and colliders of the vehicle}
        \label{fig:three graphs}
\end{figure}


\subsection{Track Design | Checkpoints}
The right and left lanes are full of checkpoints, which are invisible non-rigid colliders \cite{UnityC} shown in Figure A.2, which guide vehicles through rewards to learn that they should drive on the assigned lane unless to avoid crashing with an obstacle because crashing will result in a higher penalty than to do lane change and drive on the left lane. These checkpoints in Figure A.2 are for the vehicle driving on the right lane and will be flipped for the vehicle driving on the left lane.

\begin{figure}[ht]
    \centering
    \includegraphics[scale=0.32]{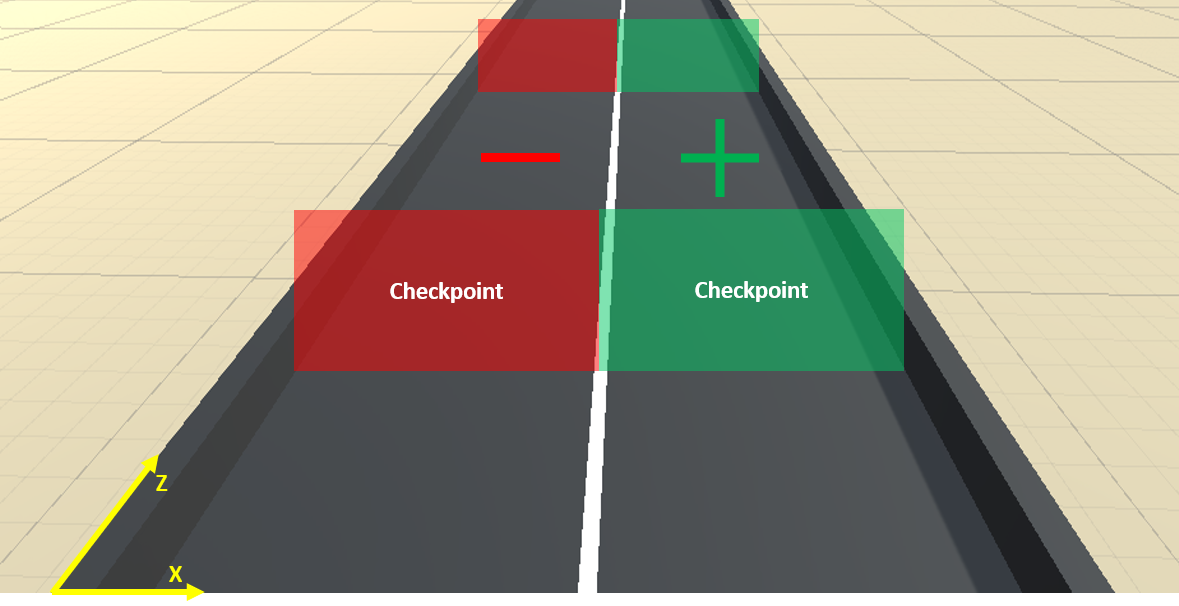}
    \caption{Checkpoints}
\end{figure}

\subsection{Experimental work}

\subsubsection{Computational Resources}
The training and testing experiments are conducted on a personal computer equipped with an Intel Core i7-9750H CPU, Nvidia GeForce RTX 2060 GPU, and 16GB of RAM. The Unity real-time 3D Game Engine \cite{Unity} is utilized as the simulator for creating the training environment and evaluating the proposed CAD control algorithm using deep RL. The training process involves the use of the ML-Agents toolkit \cite{juliani2018unity} within Unity, facilitating seamless integration between the Unity Editor and the intelligent agent training.

\subsubsection{Survey}

The trained deep RL model's performance and its compatibility with mixed traffic were verified through a survey involving 11 participants, as depicted in Figure A.3. The survey aimed to assess the driving performance achieved by the trained deep RL model for AV motion control and to conduct a mixed traffic experiment involving both AV and human drivers.

\begin{figure}[ht]
     \centering
     \begin{subfigure}[b]{0.3\textwidth}
         \centering
         \includegraphics[width=\textwidth]{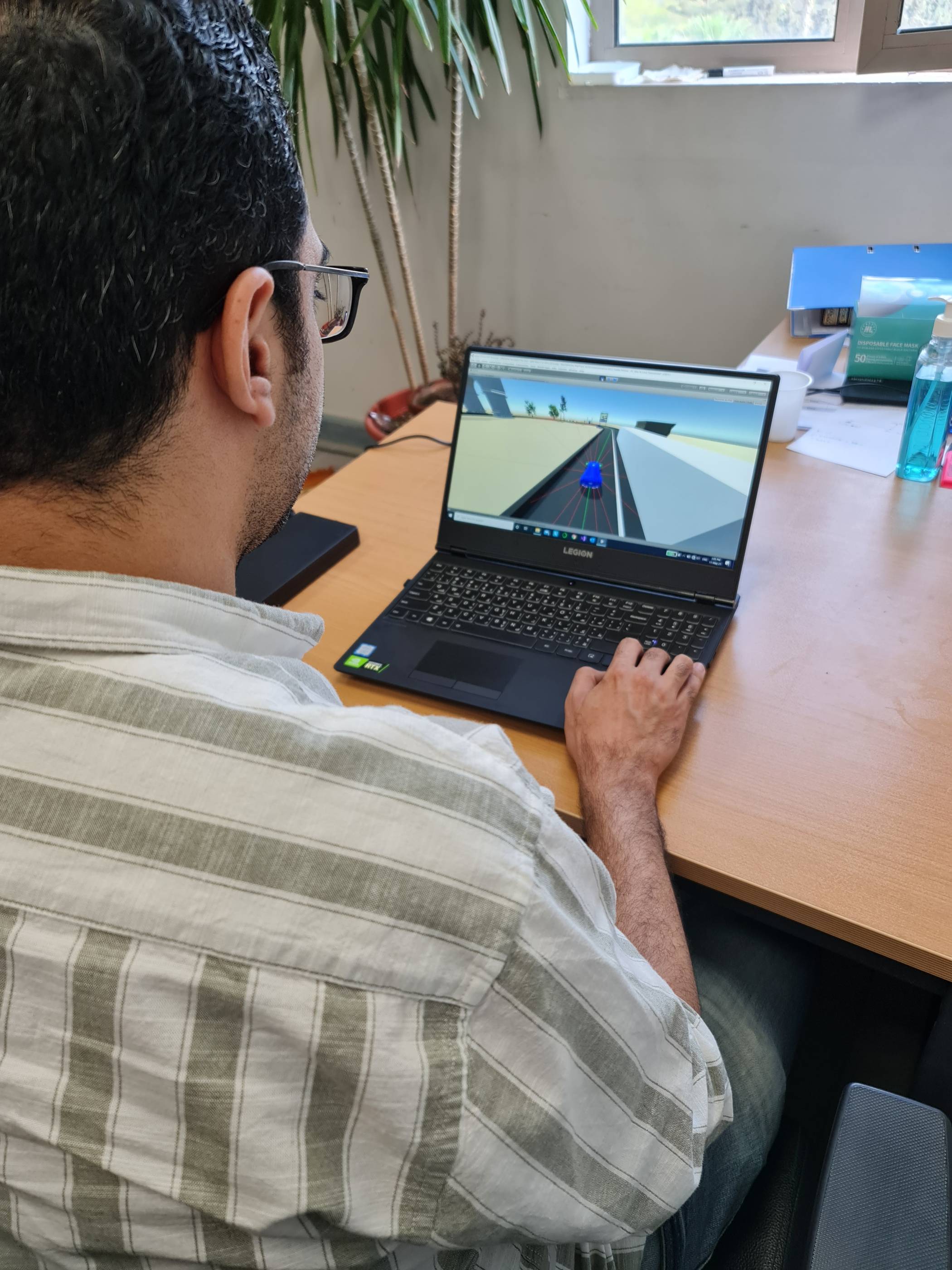}
         \label{fig:y equals x}
     \end{subfigure}
     \hfill
     \begin{subfigure}[b]{0.3\textwidth}
         \centering
         \includegraphics[width=\textwidth]{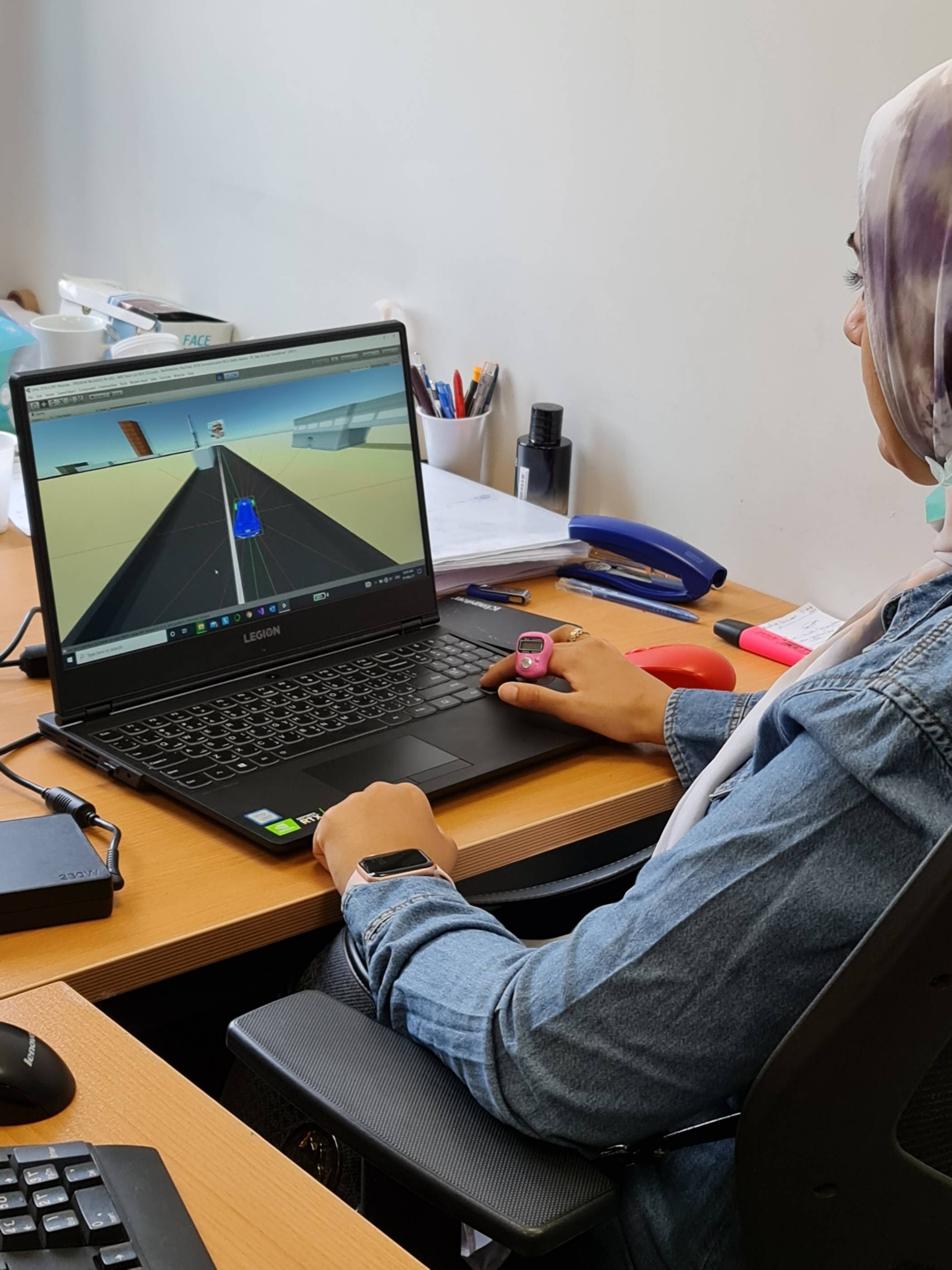}
         \label{fig:five over x}
     \end{subfigure}
     \hfill
     \begin{subfigure}[b]{0.3\textwidth}
         \centering
         \includegraphics[width=\textwidth]{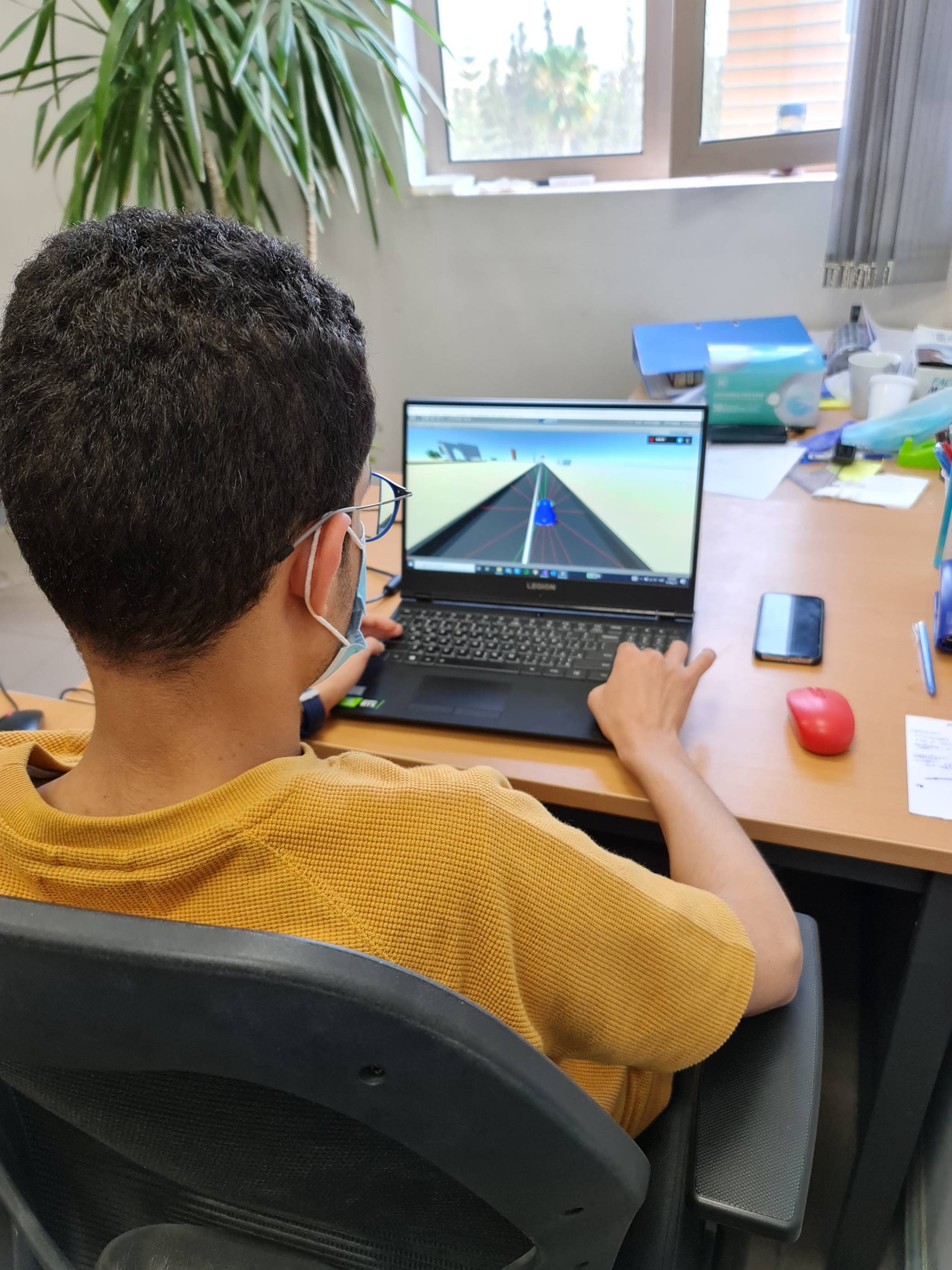}
         \label{fig:three sin x}
     \end{subfigure}
     
        \caption{Some of the participants in the survey while driving the vehicle in Unity 3D}
        \label{fig:three graphs}
\end{figure}

The 11 survey participants comprised individuals of diverse genders, backgrounds, and ages. They were introduced to the game and its mechanics. Personal information was gathered to ensure a fair and robust comparison between the participants and the automated vehicle. The following section presents the Question and Answer (Q\&A) session, providing further insights from the survey.
Participants shared their answers to the 3 questions asked prior to participating in the survey. The questions are as follows,
\begin{itemize}
    \item Q1: How long have you been driving cars?
    \item Q2: Have you ever played car games?
    \item Q3: Rate your performance in playing car games
\end{itemize}

Figure A.4 shows the answers of the participants in the survey before playing the game with the developed AV. Their answers are projected in the analysis of the obtained results.

\begin{figure}[ht]
     \centering
     \begin{subfigure}[b]{0.325\textwidth}
         \centering
         \includegraphics[width=\textwidth]{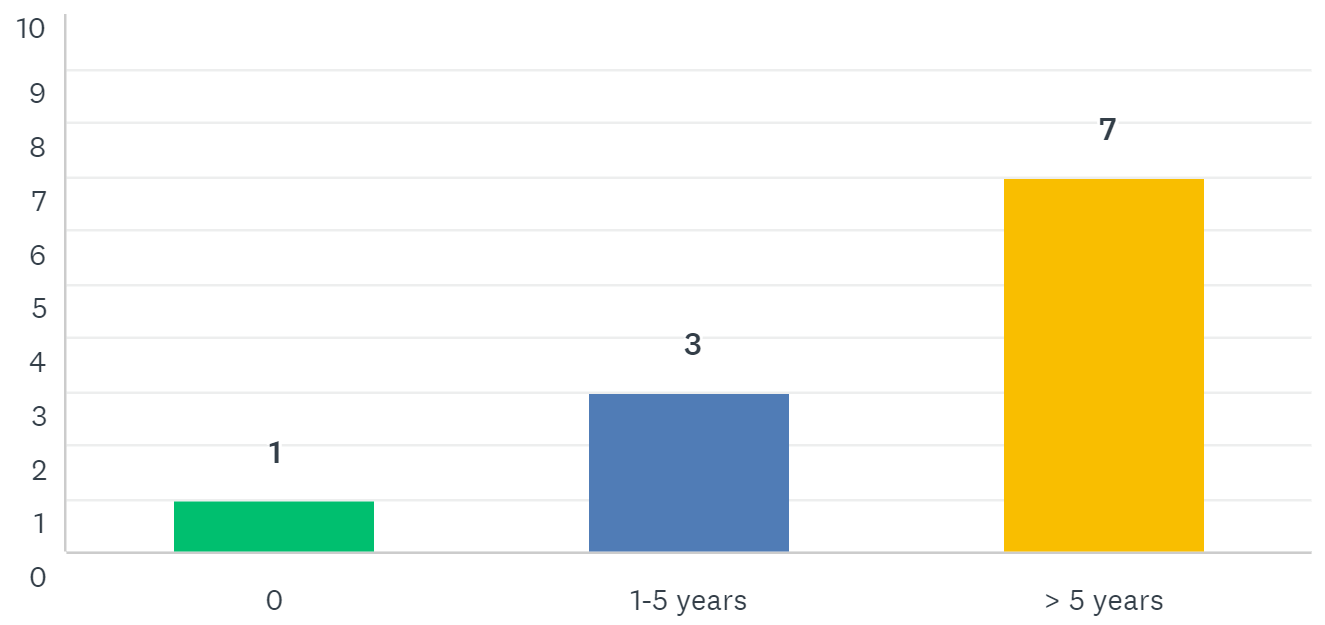}
         \caption{Answers of Q1}
     \end{subfigure}
     \begin{subfigure}[b]{0.325\textwidth}
         \centering
         \includegraphics[width=\textwidth]{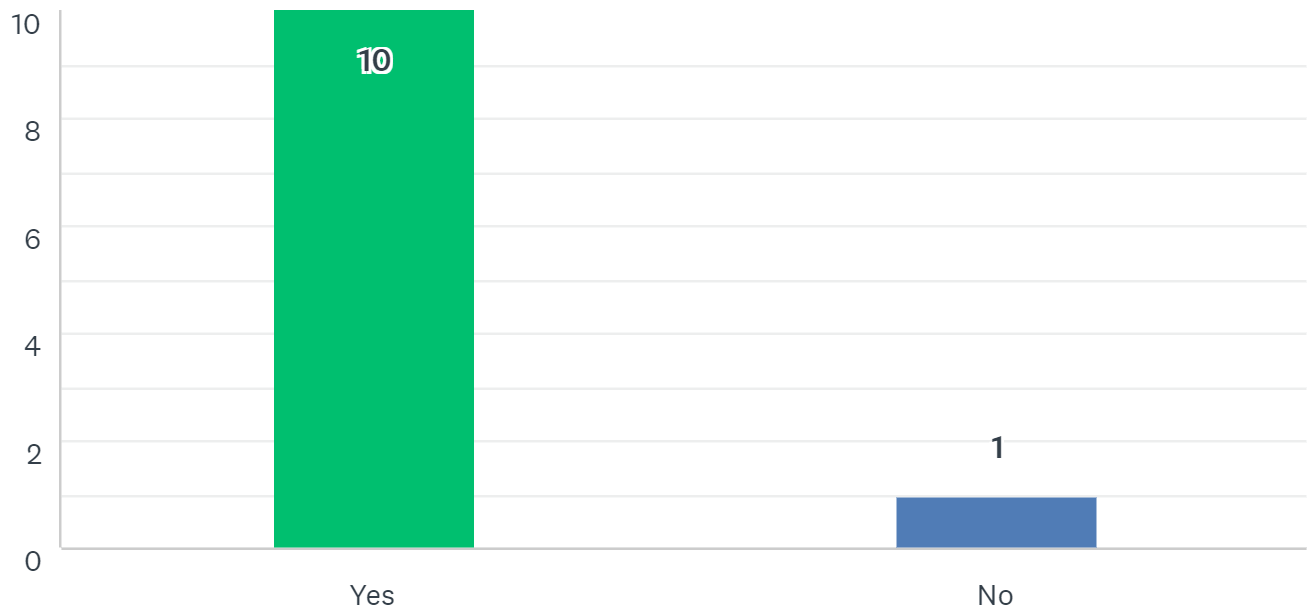}
         \caption{Answers of Q2}
     \end{subfigure}
     \begin{subfigure}[b]{0.325\textwidth}
         \centering
         \includegraphics[width=\textwidth]{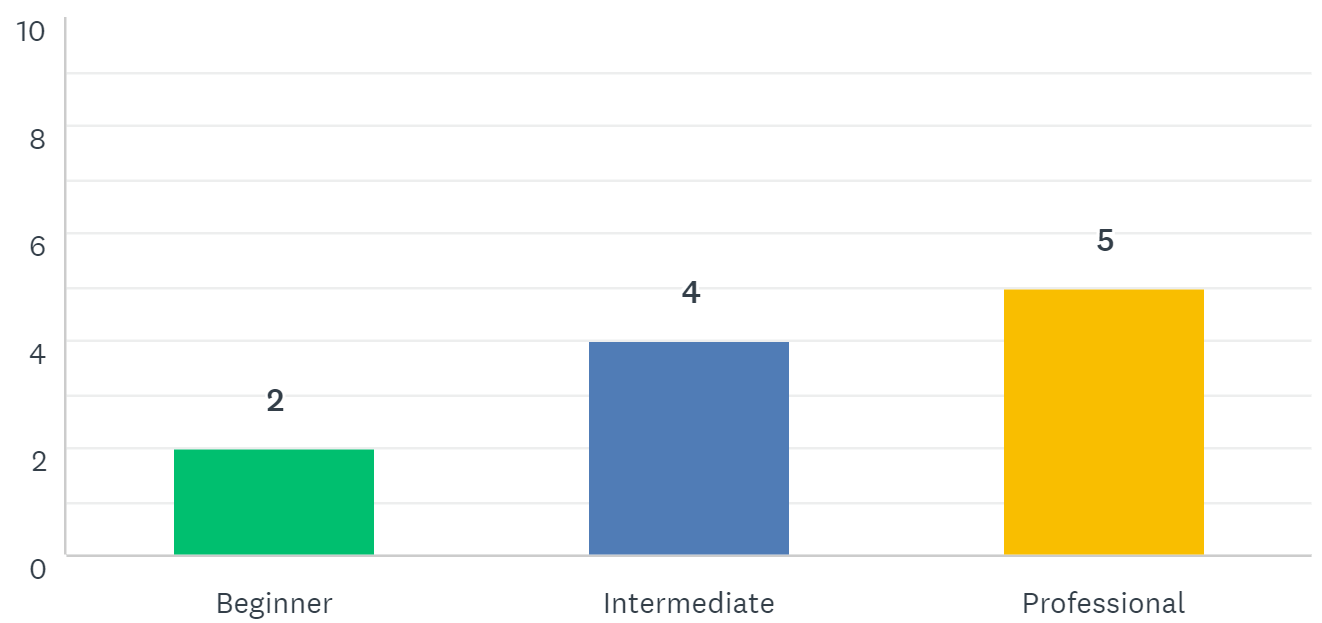}
         \caption{Answers of Q3}
     \end{subfigure}
        \caption{Summary of the participants answers on the asked questions}
        \label{fig:three graphs}
\end{figure}

\subsection{Training Curves of the Deep RL models}
Figure A.5(a) displays the training trajectory of the deep RL model used for SAV control within the urban training setting.
Figure A.5(b) illustrates the training trajectory for the red vehicle as it learns to navigate the left lane. During this phase, only the red AV is active for training, with the blue AV deactivated and V2V communication turned off.
Figure A.5(c) presents the training trajectory for the red AV as it progresses in learning to drive in the left lane. Meanwhile, the blue AV navigates the right lane. Training is exclusive to the red AV, but the blue AV shares its perception data with the red AV using a unidirectional communication topology. V2V is active for the red AV but disabled for the blue AV.
Figure A.5(d) depicts the training trajectory of the blue vehicle as it advances in learning to drive in the right lane, with the red vehicle navigating the left lane. The training is specific to the blue vehicle, yet both vehicles exchange perception data through a bidirectional communication topology. V2V is enabled for both AVs.

\begin{figure}[ht]
     \centering
     \begin{subfigure}[b]{0.49\textwidth}
         \centering
         \includegraphics[width=\textwidth]{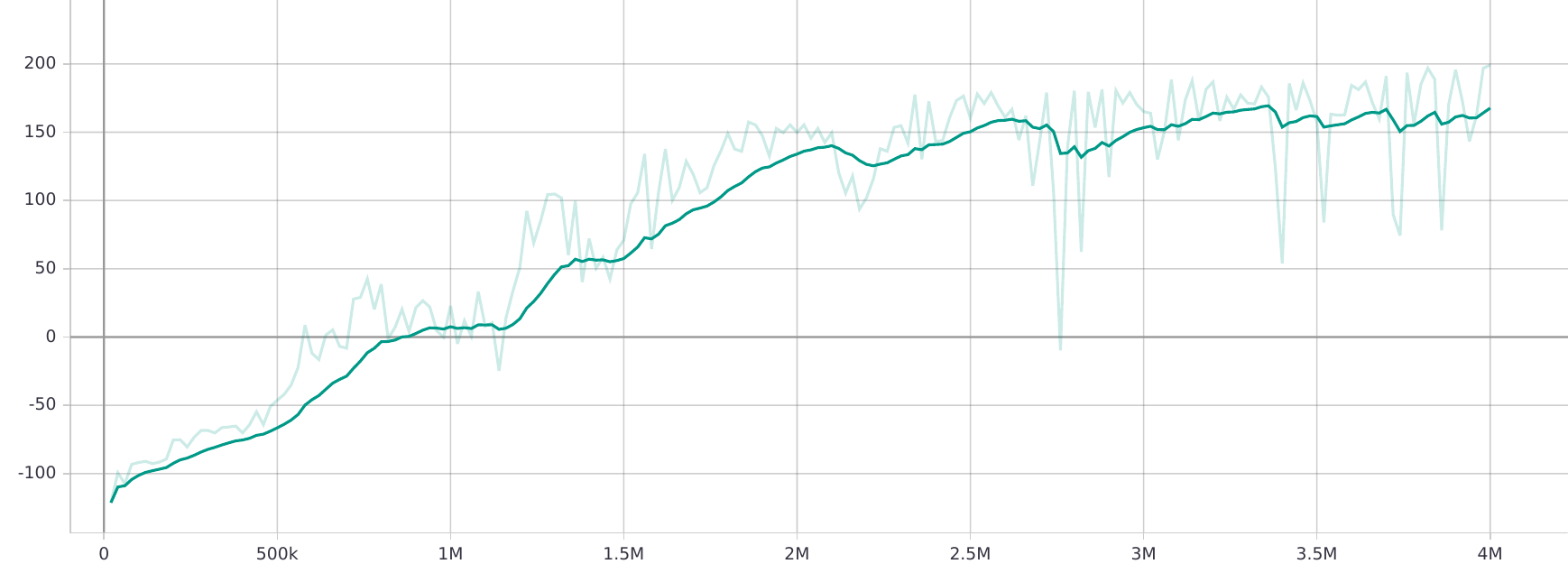}
         \caption{Stage 1 (Blue AV)}
     \end{subfigure}
     \hfill
     \begin{subfigure}[b]{0.49\textwidth}
         \centering
         \includegraphics[width=\textwidth]{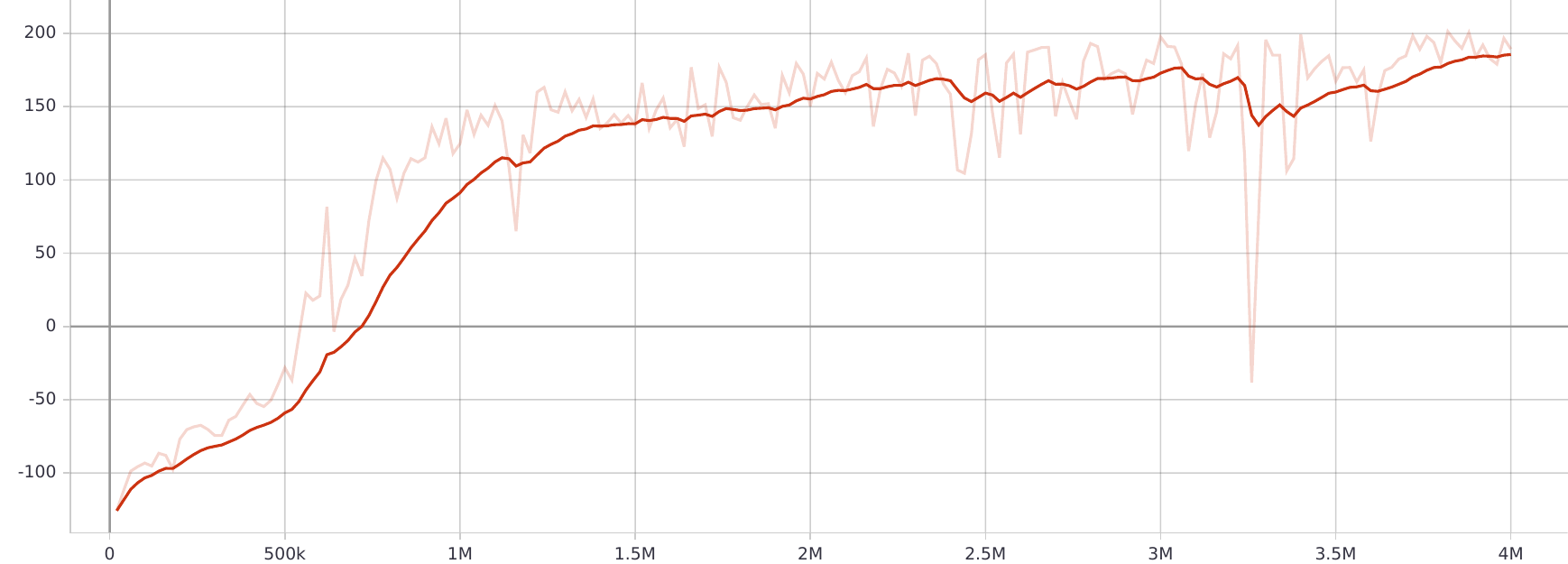}
         \caption{Stage 2 (Red AV)}
     \end{subfigure}
     \begin{subfigure}[b]{0.49\textwidth}
         \centering
         \includegraphics[width=\textwidth]{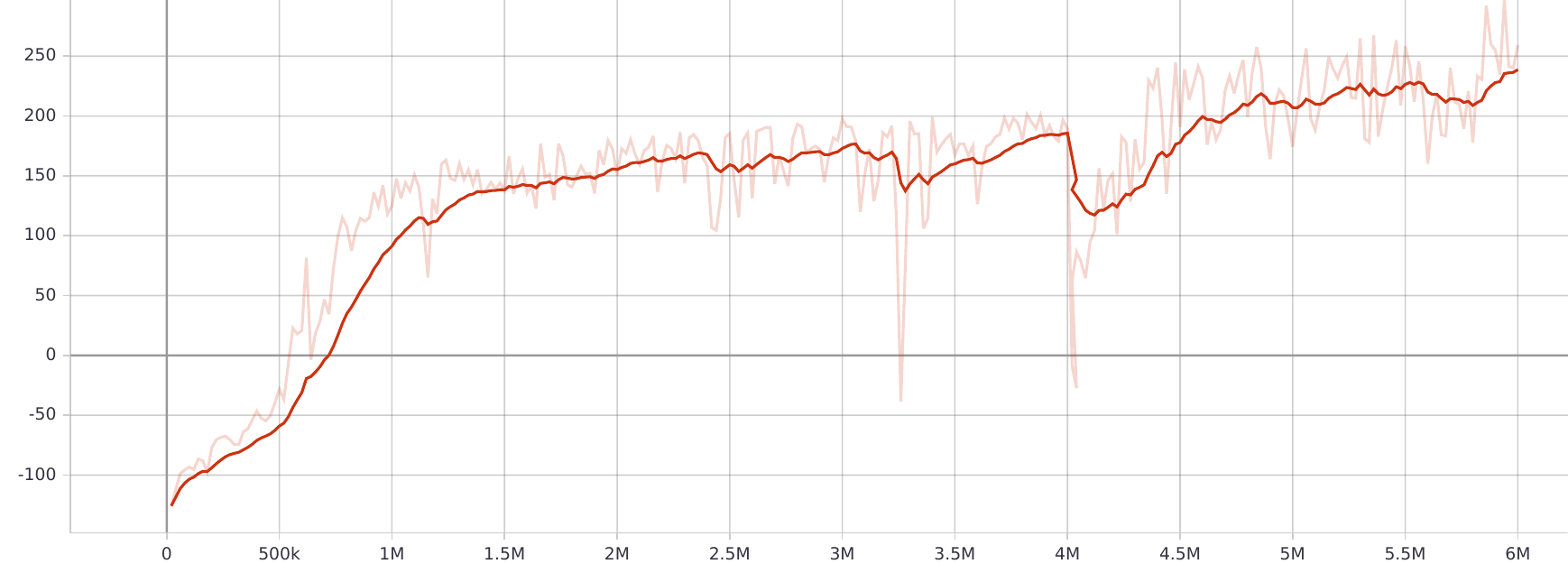}
         \caption{Stage 3 (Unidirectional V2V)}
     \end{subfigure}
     \hfill
     \begin{subfigure}[b]{0.49\textwidth}
         \centering
         \includegraphics[width=\textwidth]{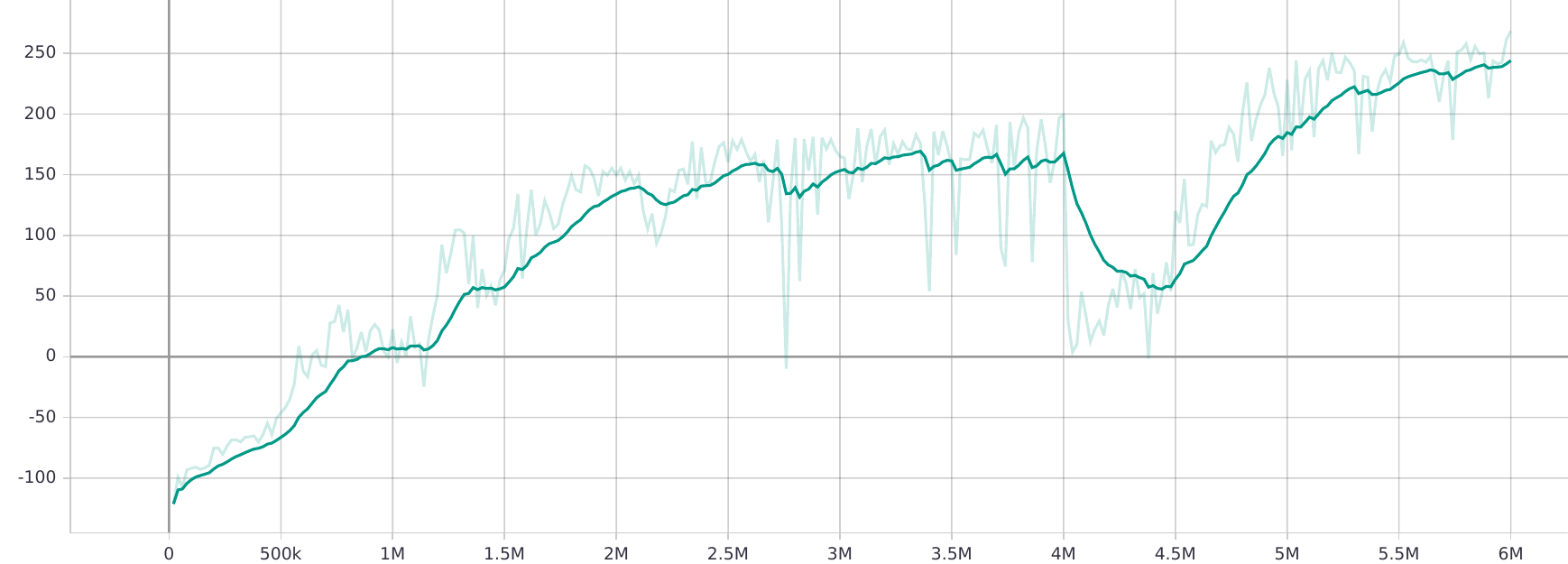}
         \caption{Stage 4 (Bidirectional V2V)}
     \end{subfigure}
        \caption{Mean cumulative reward vs training steps for the blue AV of all stages (1-4)}
        \label{fig:three graphs}
\end{figure}

\subsection{Conducted Laps Records}
Tables A.1, and A.2 show the records of the performed 10 laps on the track, demonstrating good driving with zero crashes in all laps. This reinforces the model's reliability and its capacity to navigate the track effectively under different conditions.
Table A.3 shows the average lap time for the participants and who is the winner. Participant 11 was not able to respect driving in the right lane or crash every track, which is why lap time is not considered in this experiment.

\begin{table}[ht]
\begin{minipage}{0.5\textwidth}
\centering
\captionof{table}{Blue AV lap results (Stage 1)}
\renewcommand{\arraystretch}{1.3}
\begin{tabular} {|c||c|c|} 
 \hline
 \textbf{lap no.} & \textbf{Lap Time} & \textbf{Crash}\\
\hline
lap 1 & 59.74s & No\\ 
\hline
lap 2 & 59.72s & No\\
\hline
lap 3 & 59.73s & No\\
\hline
lap 4 & 59.54s & No\\
\hline
lap 5 & 59.82s & No\\
\hline
lap 6 & 59.55s & No\\
\hline
lap 7 & 59.78s & No\\
\hline
lap 8 & 59.56s & No\\
\hline
lap 9 & 59.59s & No\\
\hline
lap 10 & 59.67s & No\\
\hline
\end{tabular}
\end{minipage}%
\begin{minipage}{0.5\textwidth}
\centering
\captionof{table}{Red AV lap results (Stage 2)}
\renewcommand{\arraystretch}{1.3}
\begin{tabular} {|c||c|c|} 
 \hline
 \textbf{lap no.} & \textbf{Lap Time} & \textbf{Crash}\\
\hline
lap 1 & 57.96s & No\\ 
\hline
lap 2 & 57.46s & No\\
\hline
lap 3 & 57.98s & No\\
\hline
lap 4 & 57.22s & No\\
\hline
lap 5 & 57.50s & No\\
\hline
lap 6 & 57.98s & No\\
\hline
lap 7 & 57.32s & No\\ 
\hline
lap 8 & 57.64s & No\\
\hline
lap 9 & 57.32s & No\\
\hline
lap 10 & 57.48s & No\\
\hline
\end{tabular}
\end{minipage}
\end{table}

\subsection{Participants Feedback}
Figure A.6 shows the participants' feedback survey after engaging in the car game with the AV. They were asked to share their opinion on the performance of the AV in comparison to their own driving abilities or human driving performance in general. A majority of the participants, approximately 64\%, believed that the AV outperformed human driving. 27\% felt that the AV's driving performance was on par with human drivers, while only one participant felt that their driving was better than the AV.

\begin{table}[ht]
\centering
\caption{Survey experiment 1 result for each lap}
\renewcommand{\arraystretch}{1.3}
\begin{tabular} {|c||c|c|c|} 
 \hline
 \textbf{Participants} & \textbf{Playing Level} & \textbf{Human Average Time} & \textbf{Winner}\\
\hline
Participant 1 & Pro & 59.39s & Human\\ 
\hline
Participant 2 & Intermediate & 60.48s & Model\\
\hline
Participant 3 & Beginner & 60.72s & Model\\
\hline
Participant 4 & Pro & 60.06s & Model\\
\hline
Participant 5 & Pro & 59.74s & Model\\
\hline
Participant 6 & Intermediate & 61.46s & Model\\
\hline
Participant 7 & Pro & 59.34s & Human\\
\hline
Participant 8 & Intermediate & 60.73s & Model\\
\hline
Participant 9 & Pro & 59.70s & Model\\
\hline
Participant 10 & Intermediate & 61.11s & Model\\
\hline
Participant 11 & Beginner & N/A & Model\\
\hline
\end{tabular}
\end{table}

\begin{figure}[ht]
    \centering
    \includegraphics[width=0.8\linewidth]{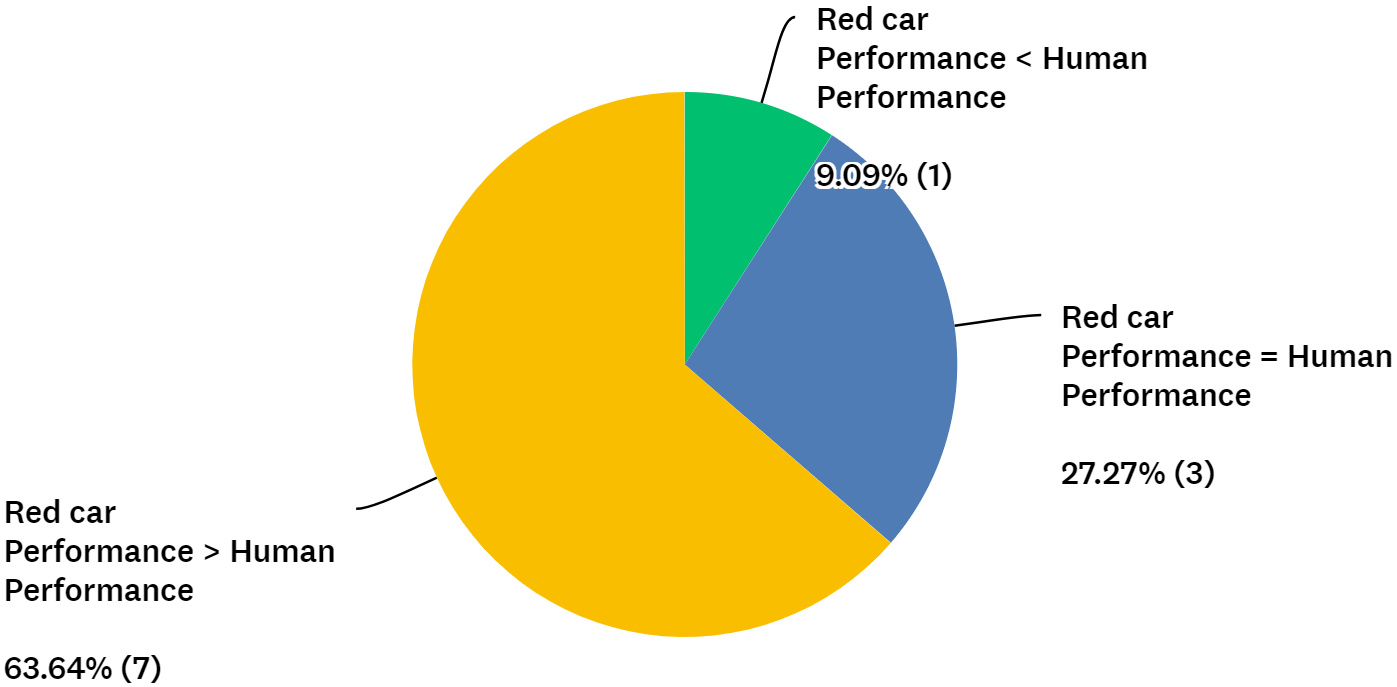}
    \caption{Rate trained model performance in driving (Red AV)}
\end{figure}

\begin{thebibliography}{10}
\providecommand{\url}[1]{#1}
\csname url@samestyle\endcsname
\providecommand{\newblock}{\relax}
\providecommand{\bibinfo}[2]{#2}
\providecommand{\BIBentrySTDinterwordspacing}{\spaceskip=0pt\relax}
\providecommand{\BIBentryALTinterwordstretchfactor}{4}
\providecommand{\BIBentryALTinterwordspacing}{\spaceskip=\fontdimen2\font plus
\BIBentryALTinterwordstretchfactor\fontdimen3\font minus \fontdimen4\font\relax}
\providecommand{\BIBforeignlanguage}[2]{{%
\expandafter\ifx\csname l@#1\endcsname\relax
\typeout{** WARNING: IEEEtran.bst: No hyphenation pattern has been}%
\typeout{** loaded for the language `#1'. Using the pattern for}%
\typeout{** the default language instead.}%
\else
\language=\csname l@#1\endcsname
\fi
#2}}
\providecommand{\BIBdecl}{\relax}
\BIBdecl

\bibitem{TaxonomyAD}
``Taxonomy and definitions for terms related to driving automation systems for on-road motor vehicles.''

\bibitem{dingus2006100}
T.~A. Dingus, S.~G. Klauer, V.~L. Neale, A.~Petersen, S.~E. Lee, J.~Sudweeks, M.~A. Perez, J.~Hankey, D.~Ramsey, S.~Gupta \emph{et~al.}, ``The 100-car naturalistic driving study, phase ii-results of the 100-car field experiment,'' United States. Department of Transportation. National Highway Traffic Safety~…, Tech. Rep., 2006.

\bibitem{choi2010crash}
E.-H. Choi, ``Crash factors in intersection-related crashes: An on-scene perspective,'' Tech. Rep., 2010.

\bibitem{outay2019v2v}
F.~Outay, F.~Kamoun, F.~Kaisser, D.~Alterri, and A.~Yasar, ``V2v and v2i communications for traffic safety and co2 emission reduction: A performance evaluation,'' \emph{Procedia Computer Science}, vol. 151, pp. 353--360, 2019.

\bibitem{van2017robust}
E.~Van~Nunen, J.~Verhaegh, E.~Silvas, E.~Semsar-Kazerooni, and N.~Van De~Wouw, ``Robust model predictive cooperative adaptive cruise control subject to v2v impairments,'' in \emph{2017 IEEE 20th International Conference on Intelligent Transportation Systems (ITSC)}.\hskip 1em plus 0.5em minus 0.4em\relax IEEE, 2017, pp. 1--8.

\bibitem{ferreira2011impact}
M.~Ferreira and P.~M. d'Orey, ``On the impact of virtual traffic lights on carbon emissions mitigation,'' \emph{IEEE Transactions on Intelligent Transportation Systems}, vol.~13, no.~1, pp. 284--295, 2011.

\bibitem{shalaby2019design}
M.~K. Shalaby, A.~Farag, O.~M. AbdelAziz, D.~M. Mahfouz, O.~M. Shehata, and E.~I. Morgan, ``Design of various dynamical-based trajectory tracking control strategies for multi-vehicle platooning problem,'' in \emph{2019 IEEE Intelligent Transportation Systems Conference (ITSC)}.\hskip 1em plus 0.5em minus 0.4em\relax IEEE, 2019, pp. 1631--1637.

\bibitem{chu2019model}
T.~Chu and U.~Kalabi{\'c}, ``Model-based deep reinforcement learning for cacc in mixed-autonomy vehicle platoon,'' in \emph{2019 IEEE 58th Conference on Decision and Control (CDC)}.\hskip 1em plus 0.5em minus 0.4em\relax IEEE, 2019, pp. 4079--4084.

\bibitem{faragreinforcement}
A.~Farag, O.~M. AbdelAziz, A.~Hussein, and O.~M. Shehata, ``Reinforcement learning based approach for multi-vehicle platooning problem with nonlinear dynamic behavior.''

\bibitem{jaritz2018end}
M.~Jaritz, R.~De~Charette, M.~Toromanoff, E.~Perot, and F.~Nashashibi, ``End-to-end race driving with deep reinforcement learning,'' in \emph{2018 IEEE International Conference on Robotics and Automation (ICRA)}.\hskip 1em plus 0.5em minus 0.4em\relax IEEE, 2018, pp. 2070--2075.

\bibitem{cai2020high}
P.~Cai, X.~Mei, L.~Tai, Y.~Sun, and M.~Liu, ``High-speed autonomous drifting with deep reinforcement learning,'' \emph{IEEE Robotics and Automation Letters}, vol.~5, no.~2, pp. 1247--1254, 2020.

\bibitem{fuchs2021super}
F.~Fuchs, Y.~Song, E.~Kaufmann, D.~Scaramuzza, and P.~D{\"u}rr, ``Super-human performance in gran turismo sport using deep reinforcement learning,'' \emph{IEEE Robotics and Automation Letters}, vol.~6, no.~3, pp. 4257--4264, 2021.

\bibitem{yang2021safety}
C.~D. Yang and D.~L. Fisher, ``Safety impacts and benefits of connected and automated vehicles: How real are they?'' pp. 135--138, 2021.

\bibitem{aradi2020survey}
S.~Aradi, ``Survey of deep reinforcement learning for motion planning of autonomous vehicles,'' \emph{IEEE Transactions on Intelligent Transportation Systems}, 2020.

\bibitem{ozioko2022road}
E.~F. Ozioko, J.~Kunkel, and F.~Stahl, ``Road intersection coordination scheme for mixed traffic (human driven and driver-less vehicles): A systematic review,'' in \emph{Science and Information Conference}.\hskip 1em plus 0.5em minus 0.4em\relax Springer, 2022, pp. 67--94.

\bibitem{sutton2018reinforcement}
R.~S. Sutton and A.~G. Barto, \emph{Reinforcement learning: An introduction}.\hskip 1em plus 0.5em minus 0.4em\relax MIT press, 2018.

\bibitem{juliani2018unity}
A.~Juliani, V.-P. Berges, E.~Teng, A.~Cohen, J.~Harper, C.~Elion, C.~Goy, Y.~Gao, H.~Henry, M.~Mattar \emph{et~al.}, ``Unity: A general platform for intelligent agents,'' \emph{arXiv preprint arXiv:1809.02627}, 2018.

\bibitem{ye2020automated}
F.~Ye, X.~Cheng, P.~Wang, C.-Y. Chan, and J.~Zhang, ``Automated lane change strategy using proximal policy optimization-based deep reinforcement learning,'' in \emph{2020 IEEE Intelligent Vehicles Symposium (IV)}.\hskip 1em plus 0.5em minus 0.4em\relax IEEE, 2020, pp. 1746--1752.

\bibitem{gupta2021embodied}
A.~Gupta, S.~Savarese, S.~Ganguli, and L.~Fei-Fei, ``Embodied intelligence via learning and evolution,'' \emph{arXiv preprint arXiv:2102.02202}, 2021.

\bibitem{aashto2001policy}
A.~AASHTO, ``Policy on geometric design of highways and streets,'' \emph{American Association of State Highway and Transportation Officials, Washington, DC}, vol.~1, no. 990, p. 158, 2001.

\bibitem{FHWA_lane}
F.~H.~S. Programs, ``Lane width,'' \url{https://safety.fhwa.dot.gov/geometric/pubs/mitigationstrategies/chapter3/3_lanewidth.cfm}, 2014.

\bibitem{tan2018survey}
C.~Tan, F.~Sun, T.~Kong, W.~Zhang, C.~Yang, and C.~Liu, ``A survey on deep transfer learning,'' in \emph{Artificial Neural Networks and Machine Learning--ICANN 2018: 27th International Conference on Artificial Neural Networks, Rhodes, Greece, October 4-7, 2018, Proceedings, Part III 27}.\hskip 1em plus 0.5em minus 0.4em\relax Springer, 2018, pp. 270--279.

\bibitem{UnitySA}
\BIBentryALTinterwordspacing
U.~Technologies. (2020) Standard assets (for unity 2018.4). [Online]. Available: \url{https://assetstore.unity.com/packages/essentials/asset-packs/standard-assets-for-unity-2018-4-32351#content}
\BIBentrySTDinterwordspacing

\bibitem{UnityC}
\BIBentryALTinterwordspacing
------. (2020) Colliders. [Online]. Available: \url{https://docs.unity3d.com/Manual/CollidersOverview.html}
\BIBentrySTDinterwordspacing

\bibitem{UnityWC}
\BIBentryALTinterwordspacing
------. (2020) Wheel collider. [Online]. Available: \url{https://docs.unity3d.com/Manual/class-WheelCollider.html}
\BIBentrySTDinterwordspacing

\bibitem{Unity}
\BIBentryALTinterwordspacing
------. (2021) Unity—manual: Execution order of event functions. [Online]. Available: \url{https: //docs.unity3d.com/Manual/executionOrder.html}
\BIBentrySTDinterwordspacing

\end{thebibliography}
\end{document}